\documentclass[a4paper,11pt]{article}
\pdfoutput=1
\usepackage{jcappub}
\usepackage{mathtools}
\usepackage{ulem}
\usepackage{bm}
\usepackage{xcolor}

\renewcommand{\thesection}{\arabic{section}}

\def\laq{~\raise 0.4ex\hbox{$<$}\kern -0.8em\lower 0.62ex\hbox{$\sim$}~}
\def\gaq{~\raise 0.4ex\hbox{$>$}\kern -0.7em\lower 0.62ex\hbox{$\sim$}~}

\def\beq{\begin{equation}}
\def\eeq{\end{equation}}
\def\bea{\begin{eqnarray}}
\def\eea{\end{eqnarray}}
\def\bean{\begin{eqnarray*}}
\def\eean{\end{eqnarray*}}

\def\laq{~\raise 0.4ex\hbox{$<$}\kern -0.8em\lower 0.62ex\hbox{$\sim$}~}
\def\gaq{~\raise 0.4ex\hbox{$>$}\kern -0.7em\lower 0.62ex\hbox{$\sim$}~}

\def\be{\begin{equation}}
\def\ee{\end{equation}}
\def \ga {\gamma}

\def\beq{\begin{equation}}
\def\eeq{\end{equation}}
\def\bea{\begin{eqnarray}}
\def\eea{\end{eqnarray}}

\def \pa {\partial}

\newcommand{\Acal}{\mathcal A}
\newcommand{\Bcal}{\mathcal B}

\newcommand{\Ecal}{\mathcal E}
\newcommand{\Ccal}{\mathcal C}
\newcommand{\Ocal}{\mathcal O}

\newcommand{\Hcal}{\mathcal H}

\def\laq{~\raise 0.4ex\hbox{$<$}\kern -0.8em\lower 0.62ex\hbox{$\sim$}~}
\def\gaq{~\raise 0.4ex\hbox{$>$}\kern -0.7em\lower 0.62ex\hbox{$\sim$}~}

\def\beq{\begin{equation}}
\def\eeq{\end{equation}}
\def\bea{\begin{eqnarray}}
\def\eea{\end{eqnarray}}
\def\bean{\begin{eqnarray*}}
\def\eean{\end{eqnarray*}}

\def \pa {\partial}

\def \ga {\gamma}

\def \Acal {\mathcal{A}}
\def \Bcal {\mathcal{B}}
\def \Ccal {\mathcal{C}}
\def \Ocal {\mathcal{O}}
\def \Lcal {\mathcal{L}}

\def \U{\Upsilon}

\title{Non-linear general relativistic effects in the observed redshift}

\author[a]{Giuseppe Fanizza}
\author[a,b]{, Jaiyul Yoo}
\author[a,c]{and Sang Gyu Biern}
\affiliation[a]{Center for Theoretical Astrophysics and Cosmology,
Institute for Computational Science, University of Z\"urich, Winterthurerstrasse 190, CH-8057, Z\"urich, Switzerland}
\affiliation[b]{Physics Institute, University of Z\"urich, Winterthurerstrasse 190, CH-8057, Z\"urich, Switzerland}
\affiliation[c]{Optotune, Bernstrasse 388 CH-8953, Dietikon, Switzerland}

\emailAdd{gfanizza@physik.uzh.ch}
\emailAdd{jyoo@physik.uzh.ch}
\emailAdd{sgbiern@physik.uzh.ch}

\abstract{We present the second-order expression for the observed redshift,
accounting for all the relativistic effects from the light propagation
and from the frame change at the observer and the source positions.
We derive the generic gauge-transformation law that any observable
quantities should satisfy, and we verify our second-order expression
for the observed redshift by explicitly checking its gauge transformation
property. This is the first time an explicit verification is made
for the second-order calculations of observable quantities. We present
our results in popular gauge choices for easy use and discuss the origin of disagreements
in previous calculations.}

\begin{document}

\maketitle

\section{Introduction}
In the recent years the Cosmic Microwave Background (CMB) has led to the precision era in observational cosmology due to WMAP \cite{Spergel:2003cb} and Planck \cite{Ade:2013zuv} satellites. Moreover new CMB missions such as CMB-S4 \cite{Abazajian:2016yjj} have been planned to further enhance our understanding of the Universe. Also in large-scale galaxy surveys significant progress has been made in the past thanks to SDSS \cite{York:2000gk}, 2dFGRS \cite{Colless:2001gk} and BOSS \cite{Dawson:2012va}. The forthcoming era of observations will provide an unprecedented amount of data with high precision. Indeed, one of the most interesting features in galaxy surveys is that they provide three-dimensional catalogs, as opposed to two-dimensional map in CMB. To full utilize 3D informations and to understand the nature of dark energy, several new galaxy surveys are planned to be operative in the very near future such as Euclid \cite{Amendola:2012ys}, DESI \cite{Aghamousa:2016zmz} and LSST \cite{Abate:2012za}. Given the level of precision set by these surveys, the theoretical predictions must be at the same level of accuracy, such that we can analyze and interpret the data without any significant systematic errors.

Motivated by this recent trend, a considerable amount of work in the last decade has been devoted to developing cosmological perturbation theory beyond linear order (see \cite{Yoo:2014sfa,BenDayan:2012wi,Umeh:2014ana} and references therein). In this respect, several works have been done, evaluating the relevant physical observables like redshift \cite{Yoo:2014sfa,Marozzi:2014kua}, luminosity distance \cite{BenDayan:2012wi,BenDayan:2012pp,Marozzi:2014kua,Fanizza:2013doa,Umeh:2014ana} and galaxy number counts \cite{DiDio:2014lka,Yoo:2014sfa,Bertacca:2014dra,Bertacca:2014wga,DiDio:2015bua}. Indeed they will be fundamental for understanding the statistical properties of inhomogeneities in the upcoming surveys. However, the second order calculations are much more involved and challenging than the linear ones. Despite the great efforts so far, there exist disagreements among the different results in literature by several groups \cite{BenDayan:2012wi,Marozzi:2014kua,Umeh:2014ana}. Hence, before applying all these results to test cosmological models with data and draw conclusions, we have to ensure that correct theoretical predictions are at our disposal.

Among all the observables, the observed redshift is the most basic quantity we have to deal with, in order to analyze data. Moreover, the usual remapping between real space and redshift space (the truly observed one) requires that we express all the physical observables in terms of redshift itself. This is already a good reason to derive the full expression of the observed redshift beyond the linear order. Moreover, the expression for the observed redshift is also directly applicable to the CMB on large scales. The possible impact of non-linear corrections on the CMB spectra has been recently studied for the leading lensing terms from both the analytical \cite{Bohm:2016gzt,Pratten:2016dsm,Marozzi:2016uob,Lewis:2016tuj,Marozzi:2016qxl,Lewis:2017ans,Bohm:2018omn} and numerical \cite{Fabbian:2017wfp,Takahashi:2017hjr,Beck:2018wud} point of view and discussed in the view of the next generation CMB experiments \cite{Marozzi:2016und}. Non-linear calculations of the observed redshift along with others observables are important and have been already performed in literature by several collaborations. However, these results have some disagreements due to the complexity inherent in general relativistic perturbation calculations. A good way to check the correctness of the theoretical expressions is to derive them without choosing any gauge. With the full expression, their gauge transformation properties can be used to check the validity. Indeed, the physical observables can be written as diffeomorphism scalar objects under gauge transformation such that they must obey the well posed gauge-transformation rules. However, none of the evaluations beyond linear order in literature are checked. In this respect, working with generic metric provides a straightforward way to achieve this goal. Then, of course, the gauge choice can be easily made a posteriori.

Motivated by this, we present the most general expression for the observed redshift at second order without choosing any gauge and we verify its validity by checking its gauge transformation properties. Finally, we are going to compare our results with literature in different gauge choices. This paper is organized as follow. In Sec. \ref{sec:GI} we discuss the importance of the coordinates lapse in terms of the gauge invariance for the physical quantities. In Sec. \ref{sec:lapses} we derive the expressions for the coordinate lapses at the observer position for a generic physical observables. In Sec. \ref{sec:redshift} we derive the second order expression for the observed redshift without choosing any gauge condition and we verify if the obtained expression gauge-transforms as expected. In Sec. \ref{sec:popular} we present the expression in some few popular gauge conditions and compare our results with the previous calculations in literature. In Sec. \ref{sec:conclusion} we discuss our results with its implications for future surveys. In Appendix we present our notation convention and summarize useful equations in the adopted GLC gauge.

\section{Building gauge-invariant combinations for physical observables}
\label{sec:GI}
\begin{figure}
\centering
\includegraphics[scale=0.75]{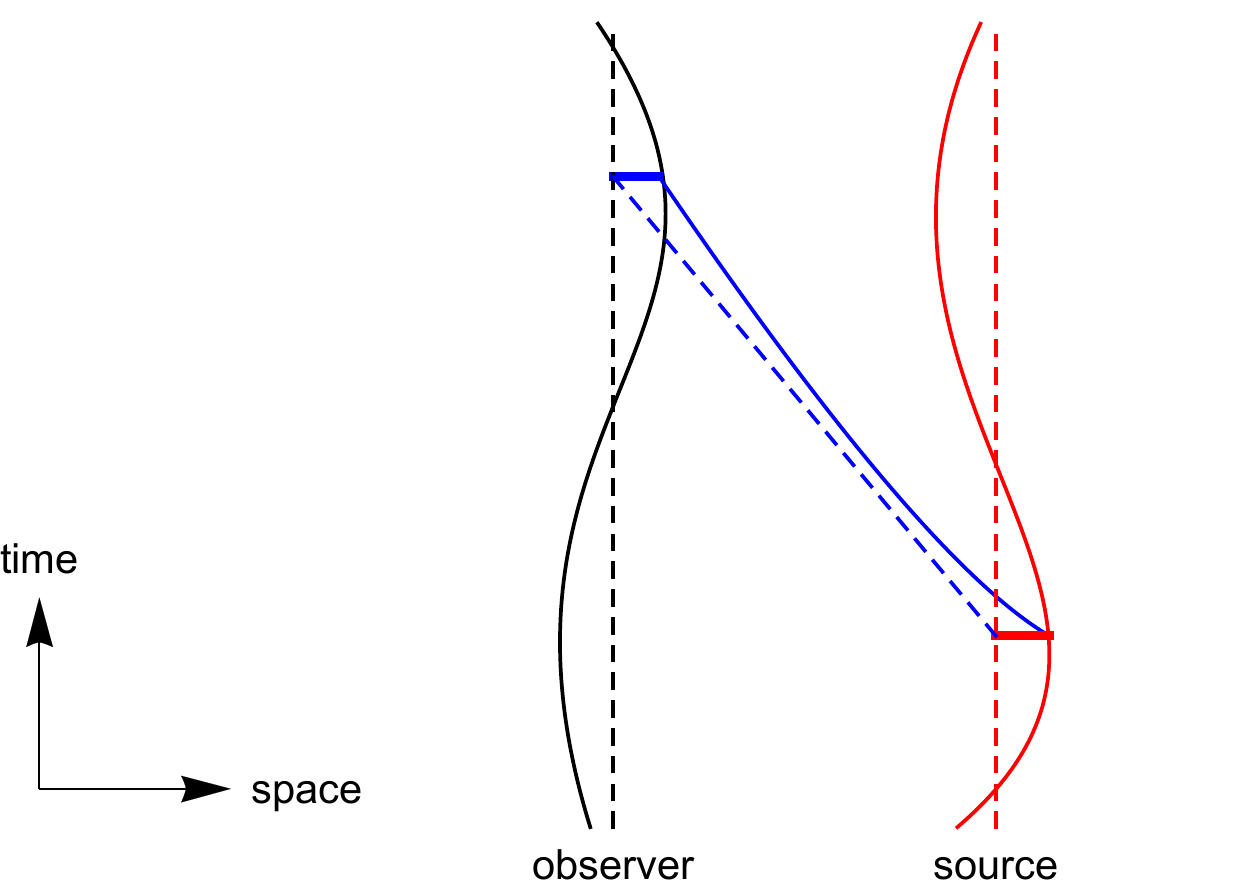}
\caption{Picture for the description of the coordinate lapse and spatial shift. Two solid curves refer to the real motions in spacetime along the geodesics for the observer $o$ (black), the source $s$ (red). The solid line connecting two geodesics represents the photon path traveling from the source to the observer (blue). Dashed lines represents the same quantities in a fiducial background. For the observer, the coordinate is aligned with the observer rest frame in the background, while the source position is defined with respect to the observed redshift and angles. As we can see, the differences between these different world-lines generate the coordinate lapses and shifts (thick lines) both at the observer and source positions, connected by the photon path.}
\label{fig:lapses}
\end{figure}
The diffeomorphism invariance is a symmetry in general relativity. Thanks to this, given a set of coordinates $x^\mu$, we can always switch to another coordinate $\tilde{x}^\mu$
\begin{equation}
\tilde{x}^\mu=\tilde{x}^\mu(x^\nu)\,,
\label{eq:diff}
\end{equation}
and the physics described with two different coordinates $x^\mu$ and $\tilde{x}^\mu$ is exactly the same at the fully non-linear level. Naturally, this underlying symmetry of the theory holds also in a perturbative description of physics. In particular, Eq. \eqref{eq:diff} can be Taylor expanded up to any desired order
\begin{equation}
\tilde{x}^\mu=x^\mu+\epsilon^\mu+\frac{1}{2}\epsilon^\nu\partial_\nu\epsilon^\mu+O(3)\,,
\label{eq:GT}
\end{equation}
where $\epsilon^\mu$ is the infinitesimal coordinate transformation and $\tilde x^\mu$ represents the same physical point in a different coordinate. According to Eq. \eqref{eq:GT}, the metric tensor $\tilde g_{\mu\nu}$ in a new set of coordinates will be different and it transforms as
\begin{equation}
\tilde{g}_{\mu\nu}(\tilde{x}^\alpha)=\frac{\partial x^\rho}{\partial \tilde{x}^\mu}\frac{\partial x^\sigma}{\partial \tilde{x}^\nu}g_{\rho\sigma}(x^\alpha)\,.
\end{equation}
After expanding to the linear order in perturbations, the two metric tensors, when evaluated at the same coordinate point $x^\mu$, are related as
\begin{equation}
\tilde{g}_{\mu\nu}=g_{\mu\nu}
-\epsilon^\rho\partial_\rho g_{\mu\nu}
-g_{\mu\rho}\partial_\nu\epsilon^\rho
-g_{\nu\rho}\partial_\mu\epsilon^\rho\,,
\label{eq:instanceN1}
\end{equation}
and this can be written in terms of the Lie derivative along $\epsilon^\mu$ as
\begin{equation}
\tilde{g}_{\mu\nu}=g_{\mu\nu}-\Lcal_{\epsilon}g_{\mu\nu}\,.
\label{eq:g_Lie}
\end{equation}
This description in terms of Lie derivative can be generalized to higher order (see Eq. \eqref{eq:Lie_derivative}) and its implication is that with the non-vanishing Lie derivative $\Lcal_{\epsilon}g_{\alpha\beta}\neq 0$, $\epsilon^\mu$ is not a Killing vector and it does not represent any fundamental symmetry or property of the physics that we want to study. The appearance of $\epsilon^\mu$ just reflects the fact that perturbations depend on how we define our background: a coordinates transformation as in Eq. \eqref{eq:GT} implies a change in the definition of our perturbations. In this respect, the appearance of $\epsilon^\mu$ in a given quantity after the transformation in Eq. \eqref{eq:g_Lie} just reflects a coordinate artifact due to the chosen reference background rather than some physically meaningful property. In other words, all the physical observables must not depend on the way we adopt to split them into the background and perturbations. This is what gauge invariance means in perturbations theory and the expressions for the physical observables should be gauge-invariant. Hence let us consider a generic physical observable described by a scalar quantity $\Ocal$ and write its gauge transformation at a given coordinate $x^\mu$. At linear order it transforms as
\begin{equation}
\tilde\Ocal(x^\mu)=\Ocal(x^\mu)-\Lcal_\epsilon\Ocal(x^\mu)\,,
\label{eq:O_GT_linear}
\end{equation}
because Eq. \eqref{eq:g_Lie} holds for any tensor quantity. It is clear that $\Ocal$ is not gauge invariant, despite being a scalar under coordinate transformations. The transformation arises because the scalar quantities $\tilde\Ocal$ and $\Ocal$ are evaluated at two different physical points but at the same coordinate value, which is the generic property of gauge transformation. We can build a gauge invariant combination by using a gauge-invariant or a
``fixed'' reference to the background position $\hat x^\mu$, such that we can express the same physical point by using the \textit{same} reference in \textit{any} coordinate system as
\begin{equation}
x^\mu=\hat x^\mu+\delta x^\mu\quad,\quad\tilde x^\mu=\hat x^\mu+\widetilde{\delta x^\mu}\,,
\label{eq:27}
\end{equation}
where $\delta x^\mu$ and $\widetilde{\delta x^\mu}$ are the perturbations in each coordinate system with respect to the fixed reference $\hat x^\mu$ and this perturbation $\delta x^\mu=\left(\delta t,\delta x^i\right)$ is called the \textit{coordinate lapse and shift}. Since $\hat x^\mu$ is coordinate-independent by construction, it has the same value in any coordinates, and the perturbations $\delta x^\mu$ and $\widetilde{\delta x^\mu}$ should be related as
\begin{equation}
\widetilde{\delta x^\mu}=\delta x^\mu+\epsilon^\mu+O(2)\,,
\label{eq:28}
\end{equation}
according to Eq. \eqref{eq:GT}. Therefore, the combination $\Ocal+\delta x^\mu\partial_\mu\Ocal$ is gauge invariant in any coordinate at first order as long as the combination is expressed at the reference coordinate $\hat x^\mu$:
\begin{equation}
\left[\widetilde{\Ocal}+\widetilde{\delta x^\mu}\partial_\mu\widetilde{\Ocal}\right]\left( \hat x^\mu \right)=\left[\Ocal+\delta x^\mu\pa_\mu\Ocal\right]\left( \hat x^\mu \right)\,.
\label{eq:linear_example}
\end{equation}
The physical interpretation of this procedure is that the extra term $\delta x^\mu\partial_\mu\Ocal$ compensates for the gauge-transformation of $\Ocal$ induced by Eq. \eqref{eq:GT} (see \cite{Yoo:2017svj} for a more detailed study)\footnote{Note that we use $\hat x^\mu$ to represent a coordinate independent reference to a physical point and it should be distinguished from a coordinate $x^\mu$. The latter $x^\mu$ changes with a coordinate transformation while our reference $\hat x^\mu$ remains unaffected.}. This results is not surprising, since we refer to a background $\hat x^\mu$ which is gauge invariant by construction. Indeed, in order to understand better this statement, we split a generic scalar into the background and perturbations and impose the diffeomorphism-invariance condition at te same physical point
\begin{equation}
\widetilde{\bar\Ocal}(\tilde x^\mu)+\widetilde{\delta\Ocal}(\tilde x^\mu)=\bar\Ocal(x^\mu)+\delta\Ocal(x^\mu)\,.
\label{eq:linear_example2}
\end{equation}
Then, by expanding around $x^\mu$, we define the gauge transformation rules for background and perturbations
\begin{equation}
\widetilde{\bar\Ocal}(x^\mu)=\bar\Ocal(x^\mu)
\quad\text{and}\quad\widetilde{\delta\Ocal}(x^\mu)=\delta\Ocal(x^\mu)-\epsilon^\mu\pa_\mu\widetilde{\bar\Ocal}( x^\mu)\,,
\end{equation}
as expected in Eq.~\eqref{eq:O_GT_linear}. Combining with Eq.~\eqref{eq:28}, we can readily
verify the gauge invariance in Eq.~\eqref{eq:linear_example}.

So far, while our argument is general, we are interested in physical
observables at physical points, for instance, the observer position
or positions along the photon path, which we denote as~$P$. That is,
we are interested in some observable quantity~$\Ocal$ evaluated at the
physical point~$P$. The coordinate value~$x^\mu_P$ will depend
on the coordinate system, and their values in two coordinate systems
are generically related as in Eq.~\eqref{eq:GT}. By introducing a fixed
reference position~$\hat x^\mu$ in 
Eq.~\eqref{eq:27}, we defined the coordinate lapse and shift associated
with the physical position~$P$. Note that while this reference
coordinate value~$\hat x^\mu_P$ is some number independent of
coordinate system, the physical position it represents differs
in each coordinate system. However, when combined with the
coordinate lapse and shift, the coordinate positions in Eq.~\eqref{eq:27}
always represent the same physical position, serving our purposes.

There are several advantages in this approach. First,
since the background reference
is independent of coordinate system, we can express and expand
our observable quantities around this point in any coordinate systems
without ambiguity. While the background quantity~$\bar \Ocal(\hat x^\mu)$
is invariant, the perturbation quantity~$\delta \Ocal(\hat x^\mu)$
gauge transforms, because the observable is evaluated not at the
physical point~$P$, but at the different point in each coordinate system.
This gauge-dependent part is compensated by the coordinate lapse and
shift as in Eq.~\eqref{eq:linear_example}, essentially by moving the coordinate-dependent physical 
position of $\hat x^\mu$ back to the physical point~$P$. So, the second advantage
is evident that by expressing all the perturbation variables 
and the coordinate lapse and shift at the reference point and 
by explicitly checking the gauge transformation property, we can
verify if our expression correctly describes the observable quantity
at the physical position.

As drawn in Fig. \ref{fig:lapses}, where we have a light-like signal emitted from a source, reaching the observer, we can apply the aforementioned procedure to both at the observer and source positions. While perturbative effects along the path are present in the expression, those perturbations along the path can always be expressed as an integration of gauge invariant variables, such that the transformation of $\delta x^\mu$ solely depends on the observer and/or the source positions, not on any point along the path. This rather surprising result is born out by Eq. \eqref{eq:O_GT_linear}: the gauge-transformation is made by a local differential operator, hence any gauge modes along the path cannot appear. As a well-known instance, this is the case for the Integrated Sachs-Wolfe effect \cite{Sachs:1967er}. Regarding the source position, its reference position $\hat x^\mu$ is usually described in terms of the observed redshift and observed angles measured by the observer. This indicates that the proper way to define the coordinate lapse and spatial shift at the source position is with respect to the fiducial background defined in terms of the observed redshift and angles (thick red line in Fig. \ref{fig:lapses}) (see \cite{BenDayan:2012pp,BenDayan:2012wi,Fanizza:2015swa}). 
Indeed the definition of the coordinate time lapse and spatial shift depends on 
our choice of a physical hypersurface we consider. In cosmology, all the observables are described in terms of observed redshift and angles, and they are
measured along the past light-cone. These observables can be used to define the physical reference point. 
Different choices can be made (see \cite{Giuseppe:2018tim}), albeit impractical.

However, the reference position of the observer is constructed in a different way as its redshift and angles are trivially zero. The physically meaningful way to define the background reference for the observer position is to use the proper time of the observer measured in its rest frame \cite{Yoo:2014vta}. In the following sections, we will construct these coordinate lapse and spatial shift at the observer and the source and use these to build gauge-invariant expressions for the physical observables.

\section{Coordinate lapse and spatial shift of the observer}
\label{sec:lapses}
In this section, we provide the analytical expression for the observer coordinate lapse and spatial shift to second order in perturbations. As described in Sec. \ref{sec:GI}, the observer position in a FLRW coordinate is different from that in the background. To compute this difference we consider the observer four-velocity
\begin{equation}
u^\mu=\frac{dx^\mu}{d\lambda}\,,
\label{eq:starting}
\end{equation}
parametrized by the proper time $\lambda$ as measured by the observer. Since the proper time $\lambda$ is also a coordinate-independent reference, the gauge transformation of the four-velocity $u^\mu$ arises due to the change in the observer path $x^\mu(\lambda)$. The difference $\delta t$ in the time coordinate is called the coordinate lapse of the observer, and the difference $\delta x^i$ in the spatial coordinate is called the (spatial) coordinate shift.

\subsection{General expression and linear order calculations}
Consider a perturbed FLRW metric given by
\begin{equation}
ds^2=-(1+2\,\mathcal{A})dt^2-2\,a\,\mathcal{B}_i\,dx^i\,dt+a^2\,\left( \delta_{ij}+2\,\mathcal{C}_{ij} \right)\,dx^idx^j\,,
\label{eq:gauge_ready}
\end{equation}
where we assumed a flat universe and chose a cartesian coordinate (see Appendix \ref{app:metric}). Given the metric in Eq. \eqref{eq:gauge_ready}, the quadri-velocity $u^\mu$ of a time-like observer, constrained by the condition $u^\mu u_\mu=-1$ can be written as
\begin{equation}
\delta u_i\equiv aV_i\qquad,\qquad u^\mu=\bar u^\mu+\delta u^\mu=\left( 1,\vec 0\right)+\left( -\Acal,\frac{\delta^{ij}}{a}\left( \Bcal_j+V_j \right) \right)\,,
\end{equation}
where the time coordinate is the cosmic time $dt$. By integrating Eq. \eqref{eq:starting} over the proper time $\lambda$ we derive the general expression for the time coordinate of the observer as\footnote{Hereafter, we use the subscript $in$ to indicate the initial condition at the early times (for instance, at the end of inflation).}
\begin{equation}
t_o-t_{in}=\int_{\lambda_{in}}^{\lambda_o}d\lambda\,
u^t(x^\mu(\lambda))
=\lambda_o-\lambda_{in}+\int_{\lambda_{in}}^{\lambda_o}d\lambda\,\delta u^t \left( x^\mu(\lambda))\right)\,.
\label{eq:derivation}
\end{equation}
This relation implies that the observer time coordinate is synchronized with the proper time in the background $\bar t_o-t_{in}=\lambda_o-\lambda_{in}$ where the proper time of the observer today in the background is
\beq
\bar t_o=\int_0^\infty\frac{dz}{(1+z)H(z)}\,.
\label{eq:35}
\eeq
The reference time coordinate $\hat t_o$ for the observer position $\hat x^\mu_o=\left( \hat t_o,\hat x^i_o \right)$ is now defined as $\hat t_o\equiv \bar t_o$. However, the inhomogeneities in the Universe affect the relation in Eqs. \eqref{eq:derivation} and the coordinate lapse $\delta t_o$ ($t_o=\bar t_o+\delta t_o$) of the observer at the linear order is
\begin{equation}
\delta t_o=\int_{\lambda_{in}}^{\lambda_o}d\lambda\,\delta u^t(x^\mu(\lambda))=-\int_{t_{in}}^{\bar t_o}dt\,\Acal\,,
\label{eq:first_time_lapses}
\end{equation}
where we changed the integration $d\lambda$ along the path of the observer to the integration $dt$ at a fixed spatial coordinate. At the linear order, the path of the observer can be approximated as a static one in computing the coordinate lapse and spatial shift. The coordinate lapse $\delta t_o$ in Eq. \eqref{eq:first_time_lapses} simply means that time coordinate of the observer is different from that in a homogeneous universe and this deviation is the cumulative time delay due to the metric perturbations along the trajectory of the observer.

Considering the geodesic motion of the observer, we can further simplify the relation for the coordinate lapse. For a generic motion, the acceleration of the observer
\begin{equation}
a_i\equiv u^\mu\nabla_\mu u_i=\pa_i\Acal+\dot{\delta u_i}\,,
\end{equation}
has to vanish, where the dot is the time derivative. The spatial velocity of the observer is related to the coordinate lapse of the observer at any time as
\begin{equation}
V_i=-\frac{1}{a}\int^{\bar t}_{t_{in}}dt\,\partial_i\Acal=\frac{1}{a}\,\partial_i\delta t\,.
\label{eq:linear_geodesic}
\end{equation}
In general, $V_i$ can be decomposed as $aV_i=\partial_i\mathcal{V}+\Omega_i$, where $\mathcal{V}$ is the velocity potential and $\Omega_i$ is a divergenceless vector. The geodesic condition dictates divergenceless vector $\Omega_i=0$, and Eq. \eqref{eq:linear_geodesic} yields $\mathcal{V}_o=\delta t_o$.
This also means that the geodesic observer has no transverse vector mode at the linear order. In the same way for the coordinate lapse, we consider the \textit{spatial coordinate shift} $\delta x_o^i$ by integrating Eq. \eqref{eq:starting} over the proper time $\lambda$
\begin{equation}
x_o^i-x_{in}^i=\int_{\lambda_{in}}^{\lambda_o}d\lambda\,\delta u^i(x^\mu(\lambda))\,.
\end{equation}
In a homogeneous universe, the observer position is $\bar x_o^i=x_{in}^i$, and it can be set $\bar x_o^i=0$ due to the spatial homogeneity. Therefore, the spatial reference $\hat x^i_o$ for the observer position coordinate is $\hat x^i_o = \bar x^i_o =0$. The spatial coordinate shift $\delta x_o^i$ at the linear order can be readily derived as
\begin{equation}
\delta x^i_o=\int_{t_{in}}^{\bar t_o}\frac{dt'}{a(t')}\,\delta^{ij}\left( \Bcal_j+V_j \right)\,.
\label{eq:spatial_lapse}
\end{equation}
While Eq. \eqref{eq:spatial_lapse} is a linear order correction, it does not appear in the observed redshift or the luminosity distance at the linear-order \cite{Biern:2016kys}. Since they are functions of time in the background, there is no spatial coordinate shift at linear order.

\subsection{Beyond the linear order}
Now we will evaluate the coordinate lapse $\delta t$ at second order. While the procedure is similar to the linear one, a subtle difference exists: at the second order, the integration over the proper time $\lambda$ can no longer be switched to an integration along the straight unperturbed geodesic, as the deviation from the straight path is also a linear order correction. These corrections are referred to as \textit{beyond the Born approximation} or \textit{post-Born effects} (see \cite{Cooray:2002mj,Krause:2009yr} for some studies about their impact on weak lensing galaxy surveys and \cite{Bohm:2016gzt,Pratten:2016dsm,Marozzi:2016uob,Lewis:2016tuj,Marozzi:2016qxl,Lewis:2017ans,Bohm:2018omn} for their consequences on CMB spectra). Keeping this in mind, we have to express $u^\mu$ in terms of second-order perturbations. Given the metric in Eq. \eqref{eq:gauge_ready} at the second order\footnote{Here and in the following, every linear perturbations will be meant to be first-order term plus second-order term. For instance $\Acal\equiv\Acal^{(1)}+\Acal^{(2)}$. On the contrary, quadratic combinations of perturbations will be considered as made of linear-order quantities so $\Bcal^i\Bcal_i\equiv \Bcal^{i(1)}\Bcal^{(1)}_i$. However, we will omit the superscript for the perturbation order to avoid clutter.}, the quadri-velocity $u^\mu$ of a time-like observer can be derived by the condition $u^\mu u_\mu=-1$ as
\begin{equation}
\bar u^\mu+\delta u^\mu=\left( 1,\vec 0\right)+\left( -\Acal+\frac{3}{2}\Acal^2-\frac{1}{2}\delta^{ij}\,\Bcal_i \Bcal_j+\frac{1}{2}\delta^{ij}V_i\,V_j,\delta u^i \right)\,,
\end{equation}
where $\delta u^i$ is determined by the geodesic equation. The general expression in Eq. \eqref{eq:derivation} for the coordinate lapse can be written at the second order as 
\begin{equation}
t_o-t_{in}=\lambda_o-\lambda_{in}+\int_{t_{in}}^{\bar t_o}d t\,\left( \delta u^t+\delta x^\nu\pa_\nu\delta u^t \right)\,,
\end{equation}
where the post-Born effects are represented by the additional term with the derivative and the deviation $\delta x^\nu$ in the path. Therefore, the second-order coordinate lapse is
\begin{equation}
\delta t_o=\int_{t_{in}}^{\bar t_o}dt\left[ \delta u^t+\delta x^\nu\left(\pa_\nu \delta t\right.\dot{\left.\right)\,\,} \right]
=\int_{t_{in}}^{\bar t_o}dt\left( \delta u^t-\delta u^\nu\pa_\nu \delta t \right)
+\left( \delta x^\nu\pa_\nu \delta t \right)_o\,,
\label{eq:second_time_lapses}
\end{equation}
where we used the linear-order relation $\delta u^\mu=\dot{\delta x^\mu}$ and performed the integration by part. Note that $\delta t_o$ in the left-hand side is the coordinate lapse of the observer at the position marked by the proper time $\lambda_o$, while its expression in the right-hand side is already expanded around the background path parametrized by the proper time $\bar t_o$. Hence we need to pay attention to the subtle difference at the second order
\begin{equation}
\delta t_o\left( \lambda \right)\equiv\delta T^{(1)}_o\left( \lambda \right)+\delta T^{(2)}_o\left( \lambda \right)+O(3)
\equiv\delta t^{(1)}_{\hat o}\left( \bar t \right)+\delta t^{(2)}_{\hat o}\left( \bar t \right)+O(3)\,,
\end{equation}
where we introduced perturbation quantities $\delta T^{(1,2)}$ and $\delta t^{(1,2)}$ by splitting $\delta t(\lambda)$ and the relation between $\delta t$ and $\delta T$ at each order is
\begin{align}
\delta t_{\hat o}^{(1)}\left( \bar t \right)=&\,\delta T_o^{(1)}\left( \lambda \right)+O(2)=\,\delta T_o^{(1)}\left( \bar t \right)+O(2)\nonumber\\
\delta t_{\hat o}^{(2)}\left( \bar t \right)=&\,\delta T_o^{(2)}\left( \lambda \right)+\delta x^\mu\pa_\mu\delta T|_\lambda+O(3)=\,\delta T_o^{(2)}\left( \bar t \right)+\delta x^\mu\pa_\mu\delta T|_{\bar t}+O(3)\,,
\label{eq:mind_the_difference}
\end{align}
where the subscript $o$ refers to quantity evaluated at $x^\mu_o$ and $\hat o$ refers to quantity evaluated at $\hat x^\mu_o$. Note that the expressions for $\delta t$ and $\delta T$ differ beyond the linear order, depending on their evaluation points, i.e. $x^\mu_o$ for $\delta T(\lambda)$ and $\hat x^\mu_o$ for $\delta t(\bar t)$, and the perturbation computation is unambiguous with evaluation at $\hat o$.
Therefore, the expression in Eq. \eqref{eq:second_time_lapses} represents the perturbation decomposition
\begin{equation}
\delta t_o\left( \lambda \right)=\delta t_{\hat o}^{(1)}\left( \bar t \right)+\delta t_{\hat o}^{(2)}\left( \bar t \right)\,.
\label{eq:A}
\end{equation}

It is interesting to notice that the integrand in right-hand side of Eq. \eqref{eq:second_time_lapses} can be rewritten as
\begin{equation}
\delta u^t-\delta u^\nu\pa_\nu \delta t=
\delta u^t
-\delta u^t\dot{\delta t}
-\delta u^i\pa_i \delta t
 =-\Acal+\frac{1}{2}\Acal^2-\frac{1}{2}\delta^{ij}\,\Bcal_i\Bcal_j+\frac{1}{2}\delta^{ij}V_i\,V_j-\delta u^i\pa_i \delta t\,.
 \label{eq:integrand}
\end{equation}
Thanks to Eq. \eqref{eq:linear_geodesic}, this means that Eq. \eqref{eq:integrand} can be written as
\begin{align}
\delta u^t-\delta u^\nu\pa_\nu \delta t=&
-\Acal+\frac{1}{2}\Acal^2-\frac{1}{2}\delta^{ij}\,\Bcal_i\Bcal_j-\frac{1}{2}\delta^{ij}V_i\,V_j-\delta^{ij}\Bcal_iV_j\,.
\label{eq:integrand_geosedic}
\end{align}
Indeed, the geodesic condition $u^\rho\nabla_\rho u_\mu=0$ at second order yields
\begin{equation}
\left( aV_i \right)\dot{}\left( 1-\Acal \right)
+\partial_i\left( \Acal-\Acal^2+\frac{1}{2}\delta^{jk}\Bcal_j\Bcal_k+\frac{1}{2}\delta^{jk}V_jV_k+\delta^{jk}\Bcal_jV_k \right)
+2\left( \Bcal^j+V^j \right)\partial_{[j} V_{i]}=0\,.
\label{eq:second_geodesic_1}
\end{equation}
Given the linear relation $\left( aV_i \right)\dot{}=-\pa_i\Acal$, we can simplify Eq. \eqref{eq:second_geodesic_1} as
\begin{equation}
\left( aV_i \right)\dot{}
+\partial_i\left( \Acal-\frac{1}{2}\Acal^2+\frac{1}{2}\delta^{jk}\Bcal_j\Bcal_k+\frac{1}{2}\delta^{jk}V_jV_k+\delta^{jk}\Bcal_jV_k \right)
+2\left( \Bcal^j+V^j \right)\partial_{[j} V_{i]}=0\,.
\label{eq:second_geodesic}
\end{equation}
As $\Omega_i=0$, the last term in Eq. \eqref{eq:second_geodesic} vanishes. In this way, the velocity field is entirely given by the velocity potential $\mathcal{V}$ up to second order and its expression is
\begin{equation}
\mathcal{V}=-\int_{t_{in}}^tdt'\left(\Acal-\frac{1}{2}\Acal^2+\frac{1}{2}\delta^{ij}\,\Bcal_i\Bcal_j
+\frac{1}{2}\delta^{ij}\partial_i\mathcal{V}\,\partial_j\mathcal{V}+\delta^{ij}\Bcal_i\partial_j\mathcal{V}\right)(t',x^i)\,.
\label{eq:velocity_geodesic}
\end{equation}
Therefore, thanks to the geodesic equation and the comparison among Eqs. \eqref{eq:second_time_lapses}, \eqref{eq:integrand_geosedic} and \eqref{eq:velocity_geodesic}, the coordinate lapse at the second order can be written in term of the velocity potential as
\begin{equation}
\delta t_o=\mathcal{V}_{\hat o}+\delta x^{\nu}_{\hat o}\pa_\nu \mathcal{V}_{\hat o}\,.
\label{eq:dtVSv}
\end{equation}
Note that the coordinate lapse for the observer is indeed valid all along the path of the observer, not just at the observer position today.

\subsection{Gauge transformation properties of the coordinate lapse and spatial shift}

Before we finish this section we check the gauge-transformation of the second-order coordinate lapse evaluated in Eq. \eqref{eq:second_time_lapses}. Let us consider the second-order gauge-transformation induced by a coordinate transformation
\begin{equation}
\tilde x^\mu=x^\mu+\epsilon^\mu
+\frac{1}{2}\epsilon^\rho\pa_\rho\epsilon^\mu\,.
\label{eq:txx}
\end{equation}
The gauge transformations for the metric components are provided in Appendix \ref{App:MT}. In particular, using Eq. \eqref{eq:non_linear_gauge_transformations}, we can easily but explicitly verify that the coordinate lapse and the spatial shift gauge transform at linear order as
\begin{align}
\widetilde{\delta t_o}(\tilde x^\mu)=&\,\delta t_o( x^\mu)+\epsilon^t_o(x^\mu)\,,\nonumber\\
\widetilde{\delta x^i_o}(\tilde x^\mu)=&\,\delta x^i_o(x^\mu)+\epsilon^i_o(x^\mu)\,.
\label{eq:GT_lienar_lapse}
\end{align}
These gauge transformation properties are in perfect agreement with what is needed to ensure the gauge-invariance of first order metric perturbations at the observer's position (see discussion after Eq. \eqref{eq:linear_example}). Similarly, at the second order, we can infer the transformation property from Eq. \eqref{eq:txx} by subtracting the reference coordinate $\hat t_o=\bar t_o$
\begin{equation}
\widetilde{\delta t}(\tilde x^\mu)=\delta t(x)+\epsilon^t(x^\mu)+\frac{1}{2}\left(\epsilon^\rho\pa_\rho\epsilon^t\right)(x^\mu)\,.
\label{eq:finalGT}
\end{equation}
Using Eq. \eqref{eq:A}, this relation is translated into
\begin{equation}
\widetilde{\delta t^{(2)}}(\hat x^\mu)=\delta t^{(2)}(\hat x^\mu)+\epsilon^t(\hat x^\mu)
+\left(\delta x^\mu\pa_\mu \epsilon^t\right)(\hat x^\mu)
+\frac{1}{2}\left(\epsilon^\mu\pa_\mu\epsilon^t\right)(\hat x^\mu)\,.
\end{equation}
Note the presence of the extra term, arising from the fact that $\epsilon^t\left( x^\mu \right)$ in Eq. \eqref{eq:finalGT} is evaluated at $x^\mu$ and we expanded it around $\hat x^\mu$. Further mind the difference between $\delta t^{(2)}(\hat x^\mu_\mu)$ and $\delta T^{(2)}(x^\mu)$ in Eq. \eqref{eq:mind_the_difference}. Now, we will explicitly verify if our expression in Eq. \eqref{eq:second_time_lapses} satisfies the above relation. First, we consider the second-order gauge transformation of the metric perturbation $\Acal^{(2)}$, given in Eq. \eqref{eq:non_linear_gauge_transformations} and the quadratic terms
\begin{align}
\frac{1}{2}\tilde\Acal^2=&\frac{1}{2}\Acal^2
+\frac{1}{2}\left(\dot{\epsilon}^t\right)^2
-\Acal\,\dot{\epsilon}^t\,,\nonumber\\
-\frac{1}{2}\tilde \Bcal^i\tilde \Bcal_i=&\,-\frac{1}{2}\Bcal^i\,\Bcal_i
-a\,\Bcal_i\,\dot{\epsilon}^i
+\frac{1}{a} \Bcal^i \partial_i \epsilon^t
-\frac{a^2}{2}\dot{\epsilon}^i\dot{\epsilon}_i
-\frac{1}{2a^2}\partial_i\epsilon^t\partial^i\epsilon^t
+\dot{\epsilon}_i\partial^i\epsilon^t\,,\nonumber\\
\frac{1}{2}\tilde V^i\tilde V_i=&\,\frac{1}{2}V^iV_i
+\frac{1}{2a^2}\partial^i\epsilon^t\partial_i\epsilon^t
+\frac{1}{a}V^i\partial_i\epsilon^t\,,\nonumber\\
-\frac{1}{a}\delta^{ij}\left( \tilde V_i+\tilde \Bcal_i \right)\pa_j \tilde{\delta t}=&\,
-\frac{1}{a}\delta^{ij}\left( V_i+ \Bcal_i \right)\pa_j \delta t
-\frac{1}{a}\delta^{ij}\left( V_i+ \Bcal_i \right)\pa_j\epsilon^t\nonumber\\
&\,-\left(\epsilon^i\pa_i \delta t\right)\dot{}
+\epsilon^i\pa_i \dot{\delta t}
-\dot\epsilon^i\pa_i\epsilon^t\,,
\end{align}
where all quantities are evaluated at the same coordinate.
Combining these altogether, we derive that the integrand term gauge-transforms as
\begin{equation}
\int_{t_{in}}^{\bar t_o}dt\left( \widetilde{\delta u^t}-\widetilde{\delta u^\nu}\pa_\nu \widetilde{\delta t} \right)
=\int_{t_{in}}^{\bar t_o}dt\left( \delta u^t-\delta u^\nu\pa_\nu \delta t \right)
+\epsilon^t_{\hat o}
-\frac{1}{2}\left(\epsilon^\rho\partial_\rho\epsilon^t\right)_{\hat o
}-\left( \epsilon^\mu \pa_\mu \delta t \right)_{\hat o}\,,
\label{eq:GTintegrand}
\end{equation}
and the remaining term at the given point gauge-transforms as
\begin{equation}
\widetilde{\delta x^\mu}\,\widetilde{\pa_\mu \delta t}=\delta x^\mu\,\pa_\mu \delta t
+\epsilon^\mu\pa_\mu \delta t
+\delta x^\mu\partial_\mu\epsilon^t
+\epsilon^\mu\partial_\mu\epsilon^t\,.
\end{equation}
Therefore, the coordinate lapse at the second-order gauge-transforms as
\begin{equation}
\widetilde{\delta t^{(2)}_{\hat o}}(\hat x^\mu)=\delta t^{(2)}_{\hat o}(\hat x^\mu)
+\epsilon^t_{\hat o}(\hat x^\mu)
+\frac{1}{2}\left(\epsilon^\rho\partial_\rho\epsilon^t\right)_{\hat o}(\hat x^\mu)
+\left( \delta x^\rho\partial_\rho\epsilon^t \right)_{\hat o}(\hat x^\mu)\,,
\label{eq:dto_G}
\end{equation}
where all the terms are evaluated at the reference coordinate $\hat x^\mu$. Let us stress that this proof is general and the geodesic equation was not invoked in the derivation, though it could simplify the derivation.

To emphasize the role of these coordinate lapse and spatial shift, we can extend the discussion in Sec. \ref{sec:GI} of the gauge-transformation properties of the observable quantities~$\Ocal$. At the second-order, a generic physical observable described by a scalar quantity $\Ocal$ gauge-transforms as
\begin{equation}
\tilde\Ocal=\Ocal-\Lcal_\epsilon\Ocal+\frac{1}{2}\,\Lcal^2_\epsilon\Ocal+O(3)\,.
\label{eq:O_second}
\end{equation}
We can then construct a gauge-invariant combination
\begin{equation}
\Ocal+\delta x^\mu\partial_\mu\Ocal+\frac{1}{2}\delta x^\mu\delta x^\nu\partial^2_{\mu\nu}\Ocal\,,
\label{eq:O_combo}
\end{equation}
provided that the coordinate lapse and spatial shift gauge-transform as
\begin{equation}
\widetilde{\delta x^\mu}(\hat x^\mu)=\delta x^\mu(\hat x^\mu)
+\epsilon^\mu(\hat x^\mu)
+\frac{1}{2}\left(\epsilon^\nu\pa_\nu\epsilon^\mu\right)(\hat x^\mu)
+\left(\delta x^\nu\pa_\nu\epsilon^\mu\right)(\hat x^\mu)\,,
\label{eq:dxo_G}
\end{equation}
where it is noted that all quantities are evaluated at the reference coordinate $\hat x^\mu$. We verify this claim by re-arranging \eqref{eq:O_second}
\begin{align}
\pa_\mu\tilde\Ocal(\hat x^\mu)=&\,\pa_\mu\Ocal-\pa_\mu\left(\Lcal_\epsilon\Ocal\right)+\frac{1}{2}\,\pa_\mu\left(\Lcal^2_\epsilon\Ocal\right)+O(3)\,,\nonumber\\
\pa^2_{\mu\nu}\tilde\Ocal(\hat x^\mu)=&\,\pa^2_{\mu\nu}\Ocal-\pa^2_{\mu\nu}\left(\Lcal_\epsilon\Ocal\right)+\frac{1}{2}\,\pa^2_{\mu\nu}\left(\Lcal^2_\epsilon\Ocal\right)+O(3)\,,
\end{align}
and by showing that the second and third terms in Eq. \eqref{eq:O_combo} gauge-transform as
\begin{align}
\widetilde{\delta x^\mu}\pa_\mu\tilde\Ocal(\hat x^\mu)=&\,\delta x^\mu\pa_\mu\Ocal
-\delta x^\mu\pa_\mu\left(\Lcal_\epsilon\Ocal\right)
-\epsilon^\mu\pa_\mu\left(\Lcal_\epsilon\Ocal\right)
+\epsilon^\mu\pa_\mu\Ocal
+\frac{1}{2}\epsilon^\nu\pa_\nu\epsilon^\mu\pa_\mu\Ocal
+\delta x^\nu\pa_\nu\epsilon^\mu\pa_\mu\Ocal\nonumber\\
=&\,\delta x^\mu\pa_\mu\Ocal
-\frac{1}{2}\Lcal_\epsilon^2\Ocal
+\Lcal_\epsilon\Ocal
-\delta x^\mu\epsilon^\nu\pa_{\mu\nu}\Ocal
-\frac{1}{2}\epsilon^\nu\epsilon^\mu\pa_{\mu\nu}\Ocal+O(3)\,,
\label{eq:term1}
\end{align}
and
\begin{align}
\frac{1}{2}\widetilde{\delta x^\mu}\widetilde{\delta x^\nu}\pa_{\mu\nu}\tilde\Ocal(\hat x^\mu)=&\,\frac{1}{2}\delta x^\mu\delta x^\nu\pa_{\mu\nu}\Ocal
+\frac{1}{2}\epsilon^\mu\epsilon^\nu\pa_{\mu\nu}\Ocal
+\delta x^\mu\epsilon^\nu\pa_{\mu\nu}\Ocal+O(3)\,.
\label{eq:term2}
\end{align}
Combining Eqs. \eqref{eq:O_second}, \eqref{eq:term1} and \eqref{eq:term2}, the combination in Eq. \eqref{eq:O_combo} is indeed gauge-invariant. This proof is completely general and can also be applied to any  generic observables. To conclude this section, we underline that Eq. \eqref{eq:O_combo} can be viewed as the second-order Taylor expansion of the observable expression $\Ocal$ around the (coordinate-independent) reference point and the condition we imposed is the diffeomorphism invariance $\tilde\Ocal(\tilde x^\mu)=\Ocal(x^\mu)$ \cite{Yoo:2017svj}.
Let us once more emphasize that the time lapse itself is not a physical quantity or gauge invariant: indeed it is defined as the difference between the 
proper time measured in the local observer frame (physical and gauge invariant) and the time coordinate in a chosen coordinate system (gauge dependent). This means that it has {\it no} any intrinsic physical meaning and hence is not directly measurable. On the other hand it must be combined with other gauge dependent terms in order to ensure the gauge invariance of the whole expression for the observable quantities.

For our practical purposes, we will use Eq. \eqref{eq:O_second} at the reference point $\hat x^\mu$, such that the full combination in Eq. \eqref{eq:O_combo} is evaluated at the fixed physical point $x^\mu_P$ that $\hat x^\mu$ refers to in any coordinate systems
\begin{equation}
\Ocal\left( x^\mu_P \right)=
\bar\Ocal(\hat x^\mu)\left[1
+\frac{\delta\Ocal}{\bar \Ocal}
+\delta x^\mu\frac{\partial_\mu\bar\Ocal}{\bar\Ocal}
+\delta x^\mu\frac{\partial_\mu\delta\Ocal}{\bar\Ocal}
+\frac{1}{2}\delta x^\mu\delta x^\nu\frac{\partial^2_{\mu\nu}\bar\Ocal}{\bar\Ocal}\right](\hat x^\mu)\,.
\label{eq:337}
\end{equation}

\section{Second-order expression for the observed redshift}
\label{sec:redshift}
In this section, we provide an explicit example of how the coordinate lapse introduced above is used to construct the second-order gauge-invariant expressions. In particular, we consider the second-order expression for the observed redshift. Since the background quantities depend only on time, the coordinate spatial shift $\delta x^i$ drops out in the linear-order description of observables quantities. Beyond the linear-order, however, the coordinate shift plays a role, as the motion of the observer deviates from the background motion, though the second-order coordinate shift still drops out in the second-order expression.

\subsection{General expressions for the observed redshift}
Let us consider the standard expression for the observed redshift defined as the ratio
\begin{equation}
1+z=\frac{\left( k^\mu u_\mu \right)_s}{\left( k^\mu u_\mu \right)_o}\equiv\frac{\Ecal(x^\mu_s)}{\Ecal(x^\mu_o)}\,,
\label{eq:redshift}
\end{equation}
where $k^\mu$ is the quadri-momentum of the photon, $u_\mu$ is quadri-velocity of the barionic fluid and $s$ and $o$ respectively represent the source and the observer positions. This relation allows us to compute the observed redshift up to any order in perturbation theory. For our purposes we split the scalar $\Ecal=k^\mu u_\mu\equiv\bar\Ecal+\delta \Ecal$ into the background and perturbations at a given position (either $x^\mu_o$ or $x^\mu_s$) and we split the observed redshift $1+z$ in the same way as
\begin{equation}
1+z=\frac{\left( \bar\Ecal
+\delta\Ecal
\right)_s}{\left( \bar\Ecal
+\delta\Ecal
\right)_o}
=\frac{\bar\Ecal_s}{\bar\Ecal_o}\left( 1
+\frac{\delta\Ecal}{\bar\Ecal}
\right)_s
\left( 1
-\frac{\delta\Ecal}{\bar\Ecal}
+\frac{\delta\Ecal^2}{\bar\Ecal^2} \right)_o
\equiv\frac{\bar\Ecal_s}{\bar\Ecal_o}\left( 1 + \Delta \right)\,,
\end{equation}
where we define the perturbation $\Delta$ as
\begin{equation}
\Delta=\left(\frac{\delta\Ecal}{\bar\Ecal}\right)_s-\left(\frac{\delta\Ecal}{\bar\Ecal}\right)_o
-\left(\frac{\delta\Ecal}{\bar\Ecal}\right)_o\left(\frac{\delta\Ecal}{\bar\Ecal}\right)_s
+\left(\frac{\delta\Ecal}{\bar\Ecal}\right)^2_o+O(3)\,.
\end{equation}
With the help of the Geodesic Light Cone gauge (see Appendix \ref{app:GLC}), we can compute the inner product $\Ecal=k^\mu u_\mu$ up to the second-order, albeit lengthy but straightforwardly,
\begin{align}
&\Delta=\,\left\{-\Acal
+\Bcal_r+V_r\right\}^s_o
-\int_{\eta_s}^{\eta_o}d\eta'\,\pa_{\eta'}\Xi
+\left\{\frac{1}{2}V_r^2
+\frac{1}{2}\bar\ga^{ab}V_aV_b
-\frac{1}{2}\,\Bcal_r^2
-\frac{1}{2}\bar\ga^{ab}\Bcal_a\,\Bcal_b
+\frac{3}{2}\,\Acal^2\right.\nonumber\\
&-\Acal\,\Bcal_r
-\Acal\,\left[ \left( \Acal
-\Ccal_{rr}
-\Bcal_r \right)_o-\int_\eta^{\eta_o}d\eta'\,\pa_{\eta'}\left( \Acal
-\Ccal_{rr}
-\Bcal_r \right) \right]
+\left[ \left( \Acal
-\Ccal_{rr}
-\Bcal_r \right)^o_\eta\right.\nonumber\\
&\left.-\int_\eta^{\eta_o}d\eta'\,\pa_{\eta'}\left( \Acal
-\Ccal_{rr}
-\Bcal_r \right) \right]\,\left( \Bcal_r+V_r \right)
+\bar\ga^{ab}\pa_a\left[ \delta w_o-\int_\eta^{\eta_o}d\eta'\,\left( \Acal
-\Ccal_{rr}
-\Bcal_r \right) \right]\,\left( \Bcal_b+V_b \right)\nonumber\\
&\left.-2\,\left( \Bcal_r+V_r \right)\Ccal_{rr}
-2\,\bar\ga^{ab}\left( \Bcal_a+V_a \right)\Ccal_{br}\right\}^s_o
-\left(V_{ro}
-\Ccal_{rro} \right)\left( V_r-\Acal
+\Bcal_r\right)^s_o\nonumber\\
&+\left( V_r
-\Ccal_{rr}+\pa_r\delta w_o \right)_o\int_{\eta_s}^{\eta_o}d\eta'\,\pa_{\eta'}\left(\Acal
-\Ccal_{rr}
-\Bcal_r\right)\,,
\label{eq:dz_standard}
\end{align}
where $\delta w_o$ is a normalization function derived in App. \ref{app:GLC} and $\Xi$ is the perturbation of the photon momentum
\begin{align}
\Xi=&\,\Acal
-\Ccal_{rr}
-\Bcal_r
-\frac{1}{2}\Acal^2
+\frac{1}{2}\Bcal^2_r
+\frac{1}{2}\bar\ga^{ab}\Bcal_a\,\Bcal_b
+2\,\Bcal_r\,\Ccal_{rr}
+2\,\bar\ga^{ab}\Bcal_a\,\Ccal_{br}
-\Acal\,\Ccal_{rr}\nonumber\\
&+2\,\bar\ga^{ab}\Ccal_{ra}\,\Ccal_{rb}
+\frac{3}{2}\,\Ccal^2_{rr}
-\left( \Acal-\Bcal_r-\Ccal_{rr} \right)^2
-\left( \Acal-\Bcal_r-\Ccal_{rr} \right)\int_\eta^{\eta_o}d\eta'\,\pa_{\eta'}\left(\Acal-\Ccal_{rr}-\Bcal_r\right)\nonumber\\
&+\left( \Acal-\Bcal_r-\Ccal_{rr} \right) \left( \Acal-\Ccal_{rr}-\Bcal_r \right)_o
+\left( \Acal-\Bcal_r-\Ccal_{rr} \right) \pa_r\delta w_o\nonumber\\
&-\bar\ga^{ab}\left( \Bcal_a+2\,\Ccal_{ar} \right)\pa_a\left[ \delta w_o-\int_\eta^{\eta_o}d\eta'\left( \Acal-\Bcal_r-\Ccal_{rr} \right) \right]\nonumber\\
&+\frac{1}{2}\bar\ga^{ab}\pa_a\left[ \delta w_o-\int_\eta^{\eta_o}d\eta'\left( \Acal-\Bcal_r-\Ccal_{rr} \right) \right]
\,\pa_b\left[ \delta w_o-\int_\eta^{\eta_o}d\eta'\left( \Acal-\Bcal_r-\Ccal_{rr} \right) \right]\nonumber\\
=&\,\Acal
-\Ccal_{rr}
-\Bcal_r
-\frac{3}{2}\Acal^2
-\frac{1}{2}\Bcal^2_r
+\frac{1}{2}\bar\ga^{ab}\Bcal_a\,\Bcal_b
+2\,\bar\ga^{ab}\Bcal_a\,\Ccal_{br}
+\Acal\,\Ccal_{rr}\nonumber\\
&+2\,\bar\ga^{ab}\Ccal_{ra}\,\Ccal_{rb}
+\frac{1}{2}\,\Ccal^2_{rr}
+2\,\Acal\,\Bcal_r
-\left( \Acal-\Bcal_r-\Ccal_{rr} \right)\int_\eta^{\eta_o}d\eta'\,\pa_{\eta'}\left(\Acal-\Ccal_{rr}-\Bcal_r\right)\nonumber\\
&+\left( \Acal-\Bcal_r-\Ccal_{rr} \right) \left( \Acal-\Ccal_{rr}-\Bcal_r \right)_o
+\left( \Acal-\Bcal_r-\Ccal_{rr} \right) \pa_r\delta w_o\nonumber\\
&-\bar\ga^{ab}\left( \Bcal_a+2\,\Ccal_{ar} \right)\pa_a\left[ \delta w_o-\int_\eta^{\eta_o}d\eta'\left( \Acal-\Bcal_r-\Ccal_{rr} \right) \right]\nonumber\\
&+\frac{1}{2}\bar\ga^{ab}\pa_a\left[ \delta w_o-\int_\eta^{\eta_o}d\eta'\left( \Acal-\Bcal_r-\Ccal_{rr} \right) \right]
\,\pa_b\left[ \delta w_o-\int_\eta^{\eta_o}d\eta'\left( \Acal-\Bcal_r-\Ccal_{rr} \right) \right]\,.
\label{eq:Delta}
\end{align}
While the second-order expression of $\Delta$ is inevitably lengthy, we recognize some pattern in Eq. \eqref{eq:dz_standard}. The first line contains the linear-order expression, but it also includes the second-order terms (e.g. $\Acal=\Acal^{(1)}+\Acal^{(2)}$) in addition to the quadratic terms. Moreover, Eqs. \eqref{eq:dz_standard} and \eqref{eq:Delta} are given in term of $k^\mu$ and $u_\mu$ as solved in term of background affine parameter, such that they already consider the expansion around the perturbed geodesic of the photon. The rest of Eq. \eqref{eq:dz_standard} accounts for the quadratic combination of the linear-order effects and the quadratic contributions in the velocity and the photon wave vectors. In Sec. \ref{sec:popular}, we will discuss the consequences of this correction at the observer position with several choices of gauge condition.

\subsection{Gauge transformation of the second-order expression}
\label{sec:gauge_redshift}
In Eq. \eqref{eq:dz_standard}, we have computed all the relativistic corrections to the observed redshift up to the second-order in perturbations. Before we further discuss their physical interpretations, let us first check the validity of the second-order expression by explicitly verifying its gauge-transformation property. First, using the equations in Appendix \ref{App:MT}, we can derive, after lengthly but straightforward manipulations,
\begin{align}
\widetilde{\delta\Ecal}=&\,\delta\Ecal
+H\,\left(\epsilon^t
-\frac{1}{2}\epsilon^\sigma\pa_\sigma\epsilon^t\right)\bar\Ecal
-\frac{1}{2}\left(\epsilon^t\right)^2\,\pa_t\left(H\,\bar{\Ecal}\right)
-\epsilon^\mu\pa_\mu\delta\Ecal\,,
\label{eq:genGauge}
\end{align}
where $H$ is the Hubble parameter and we used $\bar \Ecal \propto 1/a$. This gauge transformation of $\delta\Ecal$ is again fully consistent with what we expect for a scalar quantity under transformations from Eq. \eqref{eq:O_second}. However, we want to check the gauge-transformation of the full expression of the observed redshift and to verify that it is indeed a bi-scalar. Since the observed redshift represents the position of the source (not the observer), we can re-arrange the expression at the observer position $x^\mu_o$ in terms of the reference position $\hat x^\mu_o$ and make its perturbations gauge-invariant explicitly at the observer position. Indeed, by expanding $\Ecal(x^\mu_o)$ around the reference position $\hat x^\mu_o$
\begin{align}
\Ecal(x^\mu_o)&=\bar\Ecal\left( x^\mu_o \right)+\delta\Ecal\left( x^\mu_o \right)\nonumber\\
&=\,\left[\bar\Ecal+\delta\Ecal+\delta x^\mu\partial_\mu\bar\Ecal+\frac{1}{2}\delta x^\mu\delta x^\nu_o\partial^2_{\mu\nu}\bar\Ecal+\delta x^\mu\pa_\mu\delta\Ecal\right]\left( \hat x^\mu_o\right)\nonumber\\
&=\,\bar\Ecal_{\hat o}\left[1+\frac{\delta\Ecal}{\bar\Ecal}+\delta x^\mu\frac{\partial_\mu\bar\Ecal}{\bar\Ecal}+\frac{1}{2}\delta x^\mu\delta x^\nu\frac{\partial^2_{\mu\nu}\bar\Ecal}{\bar\Ecal}+\delta x^\mu_o\frac{\pa_\mu\delta\Ecal}{\bar\Ecal}\right]_{\hat o}\,,
\label{eq:Ecal_expansion}
\end{align}
we can readily recognize that the expression at the observer position is gauge invariant, as shown in Eq. \eqref{eq:337}. This is because $\hat x^\mu_o$ is coordinate-independent reference by construction and with $\delta x^\mu$ the whole combination in the bracket is evaluated at the physical point of the observer in any coordinate system. So the observed redshift is then further manipulated as 
\begin{align}
1+z=&\,\frac{\bar\Ecal(x^\mu_s)}{\bar\Ecal(x^\mu_o)}\Biggl[ 1+\Delta\left( x^\mu_o,x^\mu_s \right) \Biggr]\nonumber\\
=&\,\frac{\bar\Ecal(x^\mu_s)}{\bar\Ecal_{\hat o}}
\Biggl[ 1+\Delta\left( \hat x^\mu_o,x^\mu_s \right)+\delta x^\mu_o\left(\frac{\pa\Delta\left( x^\mu_o,x^\mu_s \right)}{\pa x^\mu_o}\right)_{x^\mu_o=\hat x^\mu_o} \Biggr]
\left[ 1+\delta x^\mu\frac{\pa_\mu\bar\Ecal}{\bar\Ecal}+\frac{1}{2}\delta x^\mu\delta x^\nu\frac{\pa^2_{\mu\nu}\bar\Ecal}{\bar\Ecal} \right]_{\hat o}^{-1}\nonumber\\
&\equiv\left( 1+\bar z \right)\Biggl[ 1+\delta z\left( \hat x^\mu_o,x^\mu_s \right) \Biggr]\,,
\label{eq:GI_dz}
\end{align}
where the subscript $\hat o$ represents $\hat x^\mu_o$ and distortion in the observed redshift is defined as
\begin{align}
\delta z(\hat x^\mu_o,x^\mu_s)=&\,\Delta\left( \hat x^\mu_o,x^\mu_s \right)+\delta x^\mu_{\hat o} \left(\frac{\pa\Delta\left( x^\mu_o,x^\mu_s \right)}{\pa x^\mu_o}\right)_{x^\mu_o=\hat x^\mu_o}
-\delta x^\mu_{\hat o}\left(\frac{\pa_\mu\bar\Ecal}{\bar\Ecal}\right)_{\hat o}\nonumber\\
&-\frac{1}{2}\delta x^\mu_{\hat o}\delta x^\nu_{\hat o}\left(\frac{\pa^2_{\mu\nu}\bar\Ecal}{\bar\Ecal}\right)_{\hat o}+\left( \delta x^\mu\frac{\pa_\mu\bar\Ecal}{\bar\Ecal} \right)_{\hat o}^2-\Delta\left( \hat x^\mu_o,x^\mu_s \right)\delta x^\mu_{\hat o}\left(\frac{\pa_\mu\bar\Ecal}{\bar\Ecal}\right)_{\hat o}\,.
\label{eq:obs_free_dz}
\end{align}
Note that the background redshift $\bar z$ of the source
\begin{equation}
1+\bar z\equiv \frac{\bar\Ecal(x^\mu_s)}{\bar\Ecal(\hat x^\mu_o)}=\frac{a(\hat t_o)}{a(t_s)}\,,
\end{equation}
is defined as the time coordinate of the source. It means that perturbations for our redshift in Eq. \eqref{eq:obs_free_dz} transforms as a scalar, rather than a bi-scalar. Indeed, by plugging in all the gauge transformations for perturbations in Eq. \eqref{eq:obs_free_dz}, we derive
\begin{align}
\widetilde{\delta z}=&\delta z
+\left[H\left(\epsilon^t
-\frac{1}{2}\epsilon^\sigma\pa_\sigma\epsilon^t\right)
-\frac{1}{2}a\Biggl(\frac{H}{a}\Biggr.\dot{\Biggl.\Biggr)}\left(\epsilon^t\right)^2
+H\,\delta z\,\epsilon^t
-\epsilon^\mu\pa_\mu\left(\frac{\delta\Ecal}{\bar\Ecal}\right)\right]_s\,.
\end{align}
Moreover, last term involves derivatives at the source position. It means that we can manipulate it as
\begin{equation}
-\left(\epsilon^\mu\pa_\mu\right)_s\left(\frac{\delta\Ecal}{\bar\Ecal}\right)_s+\left(\epsilon^\mu\pa_\mu\right)_s\left(\frac{\delta\Ecal}{\bar\Ecal}+\delta x^\mu\frac{\partial_\mu\bar\Ecal}{\bar\Ecal}\right)_o=-\left(\epsilon^\mu\pa_\mu\delta z\right)_s
\end{equation}
because derivatives at source for an observer term is zero. This confirms once more that $\delta z$ gauge transforms just at the source position. More explicitly, because $1+z=\left(1+\bar z\right)\left(1+\delta z\right)$, we have
\begin{align}
\widetilde{\delta z}(x^\mu_s)=&\,\delta z+H\left(\epsilon^t
-\frac{1}{2}\epsilon^\sigma\pa_\sigma\epsilon^t\right)
-\frac{1}{2}a\Biggl(\frac{H}{a}\Biggr.\dot{\Biggl.\Biggr)}\left(\epsilon^t\right)^2
+H\,\delta z\,\epsilon^t
-\epsilon^\mu\pa_\mu\delta z\,.
\label{eq:dz}
\end{align}
Indeed it is expected to transform as
\begin{equation}
\left( 1+\bar z \right)\widetilde{\delta z}=\left( 1+\bar z \right)\delta z
-\Lcal_\epsilon\left[\left(1+\bar z\right)\left(1+\delta z\right)\right]
+\frac{1}{2}\Lcal_\epsilon^2\left( 1+\bar z \right)\,,
\end{equation}
from which it follows that
\begin{equation}
\widetilde{\delta z}=\delta z
-a\,\left(\epsilon^t
-\frac{1}{2}\,\epsilon^\sigma\pa_\sigma\epsilon^t\right)\dot{\left(\frac{1}{a}\right)}
+\frac{1}{2}\,a\,\left( \epsilon^t \right)^2\ddot{\left( \frac{1}{a} \right)}
-a\,\epsilon^\mu\pa_\mu\left( \frac{\delta z}{a} \right)\,,
\end{equation}
which is exactly what we get from the explicit evaluation of gauge transformation from Eq. \eqref{eq:dz}.

Given our definition of $\delta z$, let us underline that $\delta z$ does not vanish at the observer position in the limit $s\rightarrow o$
\begin{align}
\delta z\rightarrow&\, \Delta\left( \hat x^\mu_o,x^\mu_o \right)
-\delta x^\mu_{\hat o}\left(\frac{\pa_\mu\bar\Ecal}{\bar\Ecal}\right)_{\hat o}
+\delta x^\mu_{\hat o} \left(\frac{\pa\Delta\left( x^\mu_o,x^\mu_s \right)}{\pa x^\mu_o}\right)_{x^\mu_o=\hat x^\mu_o\,,\, x^\mu_s=x^\mu_o}
\nonumber\\
&-\frac{1}{2}\delta x^\mu_{\hat o}\delta x^\nu_{\hat o}\left(\frac{\pa^2_{\mu\nu}\bar\Ecal}{\bar\Ecal}\right)_{\hat o}+\left( \delta x^\mu\frac{\pa_\mu\bar\Ecal}{\bar\Ecal} \right)_{\hat o}^2-\Delta\left( \hat x^\mu_o,x^\mu_o \right)\delta x^\mu_{\hat o}\left(\frac{\pa_\mu\bar\Ecal}{\bar\Ecal}\right)_{\hat o}\,.
\label{eq:limit}
\end{align}
The reason for this can be explained in this way: even when source and observer share the same position, there exists a difference $\delta x^\mu_{\hat o}$ between this position $x^\mu_o$ and the reference position $\hat x^\mu_o$. Hence, Eq. \eqref{eq:limit} represents nothing but the contribution $\delta z(\hat x^\mu_o,x^\mu_o)$ due to coordinate lapse and shift at the observer position. Note that the observed redshift vanishes in this limit
\begin{equation}
1+z=\left( 1+\bar z \right)\left( 1+\delta z \right)=1\,.
\end{equation}

\subsection{Observed redshift and its gauge invariance}
\label{subsec:41}
Given the full expression of the observed redshift and its gauge-transformation property, we check the consistency of our expression by deriving the coordinate lapse at the source position. The observed redshift is indeed expressed at the source position and we can express the source position by using the coordinate independent reference point $\hat x^\mu_z$ for the source position
\begin{equation}
x^\mu_s\equiv \hat x^\mu_z+\delta x^\mu_s\,,
\end{equation}
where the reference point is naturally defined in terms of the observed redshift and observed angle as
\begin{equation}
1+z=\frac{1}{a(\hat t_z)}\quad,\quad\hat x^i_z=\hat r_z\,n^i\,,
\label{eq:419}
\end{equation}
where $\hat t_z$ is the time coordinate for the observed redshift and
\begin{equation}
\hat r_z\equiv\int^z_0\frac{dz'}{H(z')}
\end{equation}
is the comoving distance to the source as given in terms of the observed redshift.
By expanding the full expression around the reference point $\hat x^\mu_z$
\begin{align}
a(t_s)&=a(\hat t_z)+\dot a\,\delta t_z+\frac{1}{2}a\left( H^2+\dot{H} \right)\,\delta t_z^2+O(3)\,,\nonumber\\
\delta z(x_s)&=\delta z(\hat x_z)+\delta \dot{z}\delta t_z+\delta x^i_z\,\pa_i\delta z+O(3)\,,
\end{align}
the expression for the observed redshift can be re-arranged around the reference point as
\begin{align}
1+z=&\frac{1}{a(t_s)}\left( 1+\delta z \right)=\frac{1}{a(\hat t_z)}\left[ 1-H\,\delta t+H^2\,\delta t^2-\frac{1}{2}\left( H^2+\dot{H} \right)\,\delta t^2\right]_z\nonumber\\
&\times\left[ 1+\delta z+\delta \dot{z}\delta t+\delta x^i\,\pa_i\delta z \right]_z\,,
\end{align}
where all terms are evaluated now at the reference point $\hat x^\mu_z$ for the source position and we mind the difference between $\delta t_z$ and $\delta T_s$, as in Eq. \eqref{eq:mind_the_difference}. This relation provides the coordinate lapse at the source position
\begin{equation}
\delta t_z
=\,\frac{1}{H}\left[\delta z
-\frac{1}{2}\delta z^2
+\frac{1}{2}\Biggl( \frac{\delta z^2}{H} \Biggr.\dot{\Biggl.\Biggr)}
+\delta x^i\pa_i \delta z\right]_z+O(3)\,,
\label{eq:dtG}
\end{equation}
where we have substituted the linear relation $H\delta t_z=\delta z$ in the non-linear terms. The results for the spatial shift is given in App. \ref{App:MT} (see Eqs. \eqref{eq:radial_shift_source} and \eqref{eq:angular_shift_source}). Briefly, the procedure of obtaining $\delta x^i_z=\left( \delta r_z,\delta\theta^a_z \right)$ in spherical coordinate consists of identifying the observed past light-cone with the background observed quantity $\hat \eta_z+\hat r_z$ through the knowledge of perturbative expression for $w$. The latter, indeed, contains all the perturbations along the observed past light-cone, then, the condition $w=\hat\eta_z+\hat r_z=\hat\eta_o$ provides the expression of $\delta r_s$, as given in Eq. \eqref{eq:radial_shift_source}. On the other hand, the angular part of the spatial shift at the source can be obtained by requiring that the GLC angles $\tilde\theta^a$ match with the observed ones $\hat \theta_s^a$, i.e. $\tilde\theta^a=\hat\theta^a_s$:
\begin{align}
\delta r_z&=\int_{\hat\eta_z}^{\hat \eta_o}\,d\eta\,\left[ \Acal-\Ccal_{rr}-\Bcal_r \right]+\frac{\delta t_{\hat o}}{a(\hat t_o)}-\frac{\delta t_z}{a(\hat t_z)}+\delta r_{\hat o}+O(2)\nonumber\\
\delta\theta^a_z&=-\int_{\hat\eta_z}^{\hat \eta_o}d\eta\left[ \bar\ga^{ab}\pa_b\left( \int_\eta^{\hat\eta_o}\,d\eta'\,\left[ \Acal-\Ccal_{rr}-\Bcal_r \right]+\frac{\delta t_{\hat o}}{a(\hat t_o)}+\delta r_{\hat o} \right)
+\Bcal^a
+2\,\Ccal^{ra}\right]+\delta\theta^a_{\hat o}+O(2)\,,
\label{eq:dxG}
\end{align}
where the boundaries are
\begin{equation}
\hat\eta_o=\int_0^\infty\frac{dz}{H(z)}\qquad\text{and}\qquad\hat\eta_z=\hat\eta_o-\hat r_z\,.
\end{equation}
The integrated terms in $\delta r_z$ and $\delta \theta^a_z$ agree with the ones presented in \cite{Yoo:2014kpa}. The only difference is that in this paper we consider also the contribution from the spatial shift at the observer, which are needed in order to get the right gauge-transformation. Indeed, by an explicit check, we can show that the coordinate lapse and shift at the source position in Eq. \eqref{eq:dz} transform as
\begin{align}
\widetilde{\delta t_z}&=\delta t_z+\epsilon^t+\frac{1}{2}\epsilon^\rho\pa_\rho\epsilon^t
+\delta t_z\,\dot{\epsilon^t}
+\delta x^i\pa_i\epsilon^t+O\left(3\right)\nonumber\\
\widetilde{\delta r_z}&=\delta r_z+\epsilon^r+O\left(2\right)\nonumber\\
\widetilde{\delta \theta^a_z}&=\delta\theta^a_z+\epsilon^a+O\left(2\right)\,,
\label{eq:source_time_shift}
\end{align}
in full agreement with our expectation described in Sec. \ref{sec:lapses}. Let us also note that in the limit $s\rightarrow o$, we safely recover that $\delta x^i_s\rightarrow \delta x^i_o$. Beyond the linear-order, we underline that the spatial shift plays a role in the second-order coordinate lapse $\delta t_s$, even though the observables in the background depend only on time. This is because the linear-order perturbations depend on all the coordinates $x^\mu$ so they naturally induce the spatial shift at the second order. However, provided that $\widetilde{\delta x^i}=\delta x^i+\epsilon^i$, we see that Eq. \eqref{eq:source_time_shift} is consistent with Eq. \eqref{eq:dxo_G}, namely the expression of physical observables in terms of observed redshift eliminates the presence of gauge transformation at the source position. Similarly, relating physical observables in terms of the observed angles and radial distance allows to eliminate all the spatial gauge dependences.

\subsection{Summary of the main expression}
To conclude this section, we provide the full expression for the observed redshift up to second-order, which is the main result of this paper. To do this, we expand also perturbations at the source position in terms of the reference frame $\hat x^\mu_z$. This implies that our set of perturbations becomes
\begin{align}
\delta z^{(1)}(\hat x^\mu_o,x^\mu_z)+\delta z^{(2)}(\hat x^\mu_o,x^\mu_z)
=&\,\delta z^{(1)}(\hat x^\mu_o,\hat x^\mu_z)+\delta z^{(2)}(\hat x^\mu_o,\hat x^\mu_z)
+\delta x^\mu_z\left(\frac{\pa\delta z^{(1)}(\hat x^\mu_o,x^\mu_s)}{\pa x^\mu_s}\right)_{x^\mu_s=\hat x^\mu_z}\nonumber\\
\equiv&\,\delta Z^{(1)}(\hat x^\mu_o,\hat x^\mu_z)+\delta Z^{(2)}(\hat x^\mu_o,\hat x^\mu_s)\,,
\end{align}
such that
\begin{align}
\delta Z^{(1)}(\hat x^\mu_o,\hat x^\mu_z)=&\,\delta z^{(1)}(\hat x^\mu_o,\hat x^\mu_z)\nonumber\\
\delta Z^{(2)}(\hat x^\mu_o,\hat x^\mu_z)=&\,\delta z^{(2)}(\hat x^\mu_o,\hat x^\mu_z)
+\delta x^\mu_z\frac{\pa\delta Z^{(1)}(\hat x^\mu_o,\hat x^\mu_z)}{\pa \hat x^\mu_z}\,.
\end{align}
We then get that
\begin{equation}
\delta Z^{(1)}(\hat x^\mu_o,\hat x^\mu_z)=\,\left(-\Acal
+\Bcal_r+V_r \right)^z_{\hat o}
-\int_{\hat\eta_z}^{\hat\eta_o}d\eta'\,\pa_{\eta'}\left( \Acal
-\Ccal_{rr}
-\Bcal_r \right)
-\frac{\Hcal_{\hat o}}{a_{\hat o}}\delta t_{\hat o}\,,
\label{eq:dz_linear}
\end{equation}
and
\begin{align}
&\delta Z^{(2)}(\hat x^\mu_o,\hat x^\mu_z)
=\,\left\{-\Acal
+\Bcal_r+V_r
+\frac{1}{2}V_r^2
+\frac{1}{2}\bar\ga^{ab}V_aV_b\right.\nonumber\\
&\left.-\frac{1}{2}\,\Bcal_r^2
-\frac{1}{2}\bar\ga^{ab}\Bcal_a\,\Bcal_b
+\frac{3}{2}\,\Acal^2
-\Acal\,\Bcal_r
-2\,\left( \Bcal_r+V_r \right)\Ccal_{rr}
-2\,\bar\ga^{ab}\left( \Bcal_a+V_a \right)\Ccal_{br}\right\}^z_{\hat o}\nonumber\\
&+\left( \Bcal_r+V_r \right)_s\left( \Acal
-\Ccal_{rr}
-\Bcal_r \right)^{\hat o}_z
-\Acal^s_{\hat o}\,\left( \Acal
-\Ccal_{rr}
-\Bcal_r \right)_{\hat o}\nonumber\\
&-\left(V_r
-\Ccal_{rr} \right)_{\hat o}\left( \Bcal_r+V_r-\Acal\right)^z_{\hat o}
-\left( \Bcal_r+V_r -\Acal\right)^z_{\hat o}\int_{\hat\eta_z}^{\hat \eta_o}d\eta'\,\pa_{\eta'}\left( \Acal
-\Ccal_{rr}
-\Bcal_r \right)\nonumber\\
&-\left( \Bcal^a+V^a \right)_z\,\pa_a\int_{\hat\eta_z}^{\hat\eta_o}d\eta'\,\left( \Acal
-\Ccal_{rr}
-\Bcal_r \right)
-\int_{\hat\eta_z}^{\hat\eta_o}d\eta'\,\pa_{\eta'}\Biggl\{\Acal
-\Ccal_{rr}
-\Bcal_r\Biggr.\nonumber\\
&\left.-\frac{3}{2}\Acal^2
-\frac{1}{2}\Bcal^2_r
+\frac{1}{2}\bar\ga^{ab}\Bcal_a\,\Bcal_b\right.
+2\,\bar\ga^{ab}\Bcal_a\,\Ccal_{br}
+\Acal\,\Ccal_{rr}
+2\,\bar\ga^{ab}\Ccal_{ra}\,\Ccal_{rb}
+\frac{1}{2}\,\Ccal^2_{rr}\nonumber\\
&+2\,\Acal\,\Bcal_r
-\left( \Acal-\Bcal_r-\Ccal_{rr} \right)\int_{\eta'}^{\hat\eta_o}d\eta''\,\pa_{\eta''}\left(\Acal-\Ccal_{rr}-\Bcal_r\right)\nonumber\\
&-\bar\ga^{ab}\left( \Bcal_a+2\,\Ccal_{ar} \right)\pa_a\left[ \int_{\eta_{in}}^{\hat\eta_o}d\eta\left[ \frac{a(\eta)}{a(\hat\eta_o)}\Acal-\Bcal_r+V_r \right]-\int_{\eta'}^{\hat\eta_o}d\eta''\left( \Acal-\Bcal_r-\Ccal_{rr} \right) \right]\nonumber\\
&+\frac{1}{2}\bar\ga^{ab}\pa_a\left[ \int_{\eta_{in}}^{\hat\eta_o}d\eta\left[ \frac{a(\eta)}{a(\hat\eta_o)}\Acal-\Bcal_r+V_r \right]-\int_{\eta'}^{\hat\eta_o}d\eta''\left( \Acal-\Bcal_r-\Ccal_{rr} \right) \right]\nonumber\\
&\Biggl.\times\pa_b\left[ \int_{\eta_{in}}^{\hat\eta_o}d\eta\left[ \frac{a(\eta)}{a(\hat\eta_o)}\Acal-\Bcal_r+V_r \right]-\int_{\eta'}^{\hat\eta_o}d\eta''\left( \Acal-\Bcal_r-\Ccal_{rr} \right) \right]\Biggr\}\nonumber\\
&+\left( \Bcal^a+V^a \right)^z_{\hat o}\,\pa_a \int_{\eta_{in}}^{\hat\eta_o}d\eta\left[ \frac{a(\eta)}{a(\hat\eta_o)}\Acal-\Bcal_r+V_r \right]
+\delta x^\mu_{\hat o} \left(\frac{\pa\Delta\left( x^\mu_o,\hat x^\mu_z \right)}{\pa x^\mu_o}\right)_{x^\mu_o=\hat x^\mu_o}\nonumber\\
&-\delta x^\mu_{\hat o}\left(\frac{\pa_\mu\bar\Ecal}{\bar\Ecal}\right)_{\hat o}
-\frac{1}{2}\delta x^\mu_{\hat o}\delta x^\nu_{\hat o}\left(\frac{\pa^2_{\mu\nu}\bar\Ecal}{\bar\Ecal}\right)_{\hat o}+\left( \delta x^\mu\frac{\pa_\mu\bar\Ecal}{\bar\Ecal} \right)_{\hat o}^2-\Delta\left( \hat x^\mu_o,\hat x^\mu_z \right)\delta x^\mu_{\hat o}\left(\frac{\pa_\mu\bar\Ecal}{\bar\Ecal}\right)_{\hat o}\nonumber\\
&+\delta x^\mu_z\frac{\pa\delta Z^{(1)}(\hat x^\mu_o,\hat x^\mu_z)}{\pa \hat x^\mu_z}\,.
\label{eq:dz_second}
\end{align}
Among all the observer terms, let us notice the presence of corrections given by the coordinate lapse and shift $\delta t_o$ and $\delta x^i_o$ in the last six lines. These terms appear here for the first time at non linear level and we will discuss in the following sections their consequences. Moreover, for computational purposes, let us underline that $\eta_{in}$ can be set equal to 0.

\section{Computation in popular gauges and comparison with previous work}
\label{sec:popular}
The computation of the observed redshift beyond the linear order has been performed in the past, with its application to the CMB and galaxy clustering \cite{DiDio:2014lka,Yoo:2014kpa}. As we emphasized, the non-linear perturbation calculations are lengthy and complicated by nature. In Sec. \ref{sec:redshift}, we explicitly presented the most robust calculation of the observed redshift at the second-order by verifying its gauge-transformation property. To facilitate the comparison of our result, we provide its expression in a few popular gauges: the N-Body Gauge (NB), the Synchronous Gauge (SG) and the Conformal Newtonian Gauge (NG). In all of these cases, we will consider the scalar perturbations up to second order, while the vector and the tensor perturbations are considered only at second order.

\subsection{Linear N-body gauge}
First of all, we want to provide our result in the so-called linear N-body gauge \cite{Fidler:2015npa,Fidler:2016tir}. This choice is particularly helpful when we want to interpret results from N-body codes as relativistic effects. In particular, this gauge allows to write relativistic equations in the same form as the Newtonian ones, so it allows an immediate comparison with N-body simulations when radiation density and pressure are negligible. In this case, we just focus on scalars at linear order because at non-linear level this gauge choice about the relativistic equations implies more involved conditions than the linear ones. The generalization of N-body gauge to second order is a non trivial task which we are not going to face here. More specifically, starting from the generic perturbed metric
\begin{equation}
ds^2=-(1+2\,\Acal)dt^2-2\,a\,B_i\,dx^i\,dt+a^2\,\left( \delta_{ij}+2\,\Ccal_{ij} \right)\,dx^idx^j\,,
\end{equation}
N-body gauge consists of requiring that a traceless set of spatial metric perturbations and Comoving choice for temporal gauge condition, namely
\begin{equation}
\Ccal^i_{\,\,\,i}=0\quad\text{and}\quad\mathcal{V}=0\,.
\label{eq:NB_gauge}
\end{equation}
Therefore, after the decomposition where only scalars are taken into account, we get
\begin{align}
&\qquad\qquad\Bcal_r=\,\pa_r\beta\,,\qquad\qquad
\Bcal_a=\,\pa_a\beta\,,\qquad\qquad\Acal=\phi\,,\nonumber\\
\Ccal_{rr}=&\,\psi+\pa^2_r\ga\qquad\,,\qquad
\Ccal_{ra}=\,\pa^2_{ra}\ga-\frac{1}{r}\pa_a\ga\qquad,\qquad
\Ccal_{ab}=\,\psi\,\bar\ga_{ab}+\nabla_a\pa_b\ga\,,
\end{align}
where first of Eqs. \eqref{eq:NB_gauge} implies
\begin{equation}
\psi+\frac{1}{3}\Delta\ga=0\,,
\end{equation}
where $\Delta$ is the 3-dimensional Laplacian. Moreover, when radiation pressure and density are negligible, Eqs. \eqref{eq:NB_gauge} lead to the further condition $\ga'=0$. Hence let us consider the redshift perturbations just for scalars and express our results in terms of the gauge invariant variables
\begin{equation}
\phi_\chi=\,\phi-\frac{\chi'}{a}\,,\qquad\qquad
\psi_\chi=\,\psi-\frac{\Hcal}{a}\,\chi\,,\qquad\qquad
V^\chi_i=\,V_i+\frac{1}{a}\pa_i\chi\,,
\label{eq:GI_var}
\end{equation}
where $\chi\equiv a\left(\beta+\gamma'\right)$. Here, the notation indicates that the variable $X_Y$ is equal to $X$ in the gauge where $Y=0$. In this way, first of all, we can divide the coordinate time lapse into its gauge invariant and gauge dependent part
\begin{equation}
\delta t_{\hat o}=-\frac{1}{a_{\hat o}}\int_{\eta_{in}}^{\hat \eta_o}d\eta\,a\,\phi=-\frac{1}{a_{\hat o}}\int_{\eta_{in}}^{\hat \eta_o}d\eta\,a\,\phi_\chi-\chi_{\hat o}\,,
\label{eq:GI_tl}
\end{equation}
and then, after some manipulations, linear redshift can be written as
\begin{align}
&\delta Z(\hat x^\mu_o,\hat x^\mu_s)=\,\left[-\phi_\chi
+V^\chi_r
-\frac{\Hcal}{a}\chi\right]^z_{\hat o}
-\int_{\hat \eta_z}^{\hat\eta_o}d\eta\left( \phi_\chi-\psi_\chi \right)'
+\frac{\Hcal_{\hat o}}{a_{\hat o}}\delta t_{\hat o}\nonumber\\
=&\,\left[-\phi_\chi
+V^\chi_r
-\frac{\Hcal}{a}\chi\right]^z_{\hat o}
-\int_{\hat \eta_z}^{\hat\eta_o}d\eta\left( \phi_\chi-\psi_\chi \right)'
-\frac{\Hcal_{\hat o}}{a_{\hat o}}\int_{\eta_{in}}^{\eta_o}d\eta\,a\,\phi_\chi
-\frac{\Hcal_{\hat o}}{a_{\hat o}}\chi_{\hat o}\nonumber\\
=&\,\left[-\phi_\chi
+V^\chi_r\right]^z_{\hat o}
-\int_{\hat \eta_z}^{\hat\eta_o}d\eta\left( \phi_\chi-\psi_\chi \right)'
-\frac{\Hcal_{\hat o}}{a_{\hat o}}\int_{\eta_{in}}^{\eta_o}d\eta\,a\,\phi_\chi
-\frac{\Hcal_z}{a_z}\chi_z\,.
\label{eq:NB_dz}
\end{align}
This result is still general. If we now apply condition for the N-Body gauge, we get that $\chi=a\beta$ and then
\begin{align}
\delta Z_{NB}(\hat x^\mu_o,\hat x^\mu_s)
=&\,\left[-\phi_\chi
+V^\chi_r\right]^z_{\hat o}
-\int_{\hat\eta_z}^{\hat\eta_o}d\eta\left( \phi_\chi-\psi_\chi \right)'
-\frac{\Hcal_{\hat o}}{a_{\hat o}}\int_{\eta_{in}}^{\hat\eta_o}d\eta\,a\,\phi_\chi
-\Hcal_z\beta_z\,.
\label{eq:NB_dz_lapse}
\end{align}
This result can be compared with the one obtained in \cite{Adamek:2017kir}. We notice that our result looks different with respect to the one presented in \cite{Adamek:2017kir}, where the boundary term at the observer is $\left(\Hcal\beta\right)_o$. However, the two results are in quantitative agreement: indeed, by recalling Eq. \eqref{eq:dtVSv}, we get that time lapse in the N-Body gauge is 0. From Eq. \eqref{eq:GI_tl}, we then get in the N-Body gauge that
\begin{equation}
\frac{1}{a_{\hat o}}\int_{\eta_{in}}^{\hat \eta_o}d\eta\,a\,\phi_\chi=-a_{\hat o}\beta_{\hat o}\,,
\end{equation}
which restore the quantitative agreement. We underline, instead, that our result is gauge invariant at the observer, while the one presented in \cite{Adamek:2017kir} is not. This is not surprising: indeed, N-body gauge shares the same properties of the Comoving gauge as well as the linear Synchrounous gauge in the time part of metric perturbations. Then, because boundary terms at the observer are built to be gauge invariant in all the gauge choices and equal to the ones in gauges where coordinate time lapse is null, it is a natural consequence that our result is gauge invariant and at the same time in quantitative agreement with the one in \cite{Adamek:2017kir}. This linear example is illustrative for what happens also in the Synchronous gauge, where coordinate time lapse is null too.

\subsection{Synchronous Gauge}
Given the metric form in Eq. \eqref{eq:gauge_ready_app}, the Synchronous gauge is defined by fixing the scalar fields in the decomposition in Eqs. \eqref{eq:decompositions} as
\begin{equation}
ds^2=-dt^2+a^2\,\left[ \delta_{ij}+2\,\Ccal_{ij}\,dx^idx^j\right]\quad\text{with}\quad\alpha=B_i=\mathcal{V}=0\,,
\label{eq:SG}
\end{equation}
where we assume no vectors and tensors at linear-order. Note that we further impose the comoving gauge condition $\mathcal{V}=0$, which is often implicitly assumed \cite{Yoo:2014vta}. We first compute the useful quantities
\begin{align}
\Ccal_{rr}=&\,\psi+\pa^2_r\ga+\pa_rC^{(2)}_r+C^{(2)}_{rr}\,,\nonumber\\
\Ccal_{ra}=&\,\pa^2_{ra}\ga-\frac{1}{r}\pa_a\ga+\pa_{(a}C^{(2)}_{r)}-\frac{1}{r}C^{(2)}_a+C^{(2)}_{ra}\,,\nonumber\\
\Ccal_{ab}=&\,\psi\,\bar\ga_{ab}+\nabla_a\pa_b\ga+\nabla_{(a}C^{(2)}_{b)}+C^{(2)}_{ab}\,.
\label{eq:SG_dec}
\end{align}
We neglect the second-order quantities in the quadratic combination of $\Ccal_{rr}$, $\Ccal_{ra}$ or $\Ccal_{ab}$. Therefore, it is evident from Eqs. \eqref{eq:first_time_lapses} and \eqref{eq:spatial_lapse} the coordinate lapse $\delta t_o$ and spatial shift $\delta x^i_o$ vanish at the linear order in this gauge. Hence, from Eq. \eqref{eq:dz_second}, we get that
\begin{align}
&\delta Z^{(2)}_{SG}(\hat x^\mu_o,\hat x^\mu_z)=\,
-\int_{\hat\eta_z}^{\hat\eta_o}d\eta'\,\pa_{\eta'}\left\{
-\Ccal_{rr}
+\frac{1}{2}\,\Ccal^2_{rr}
+2\,\bar\ga^{ab}\Ccal_{ra}\,\Ccal_{rb}
-\Ccal_{rr}\int_{\eta'}^{\hat\eta_o}d\eta''\,\pa_{\eta''}\Ccal_{rr}\right.\nonumber\\
&\left.
+\frac{1}{2}\bar\ga^{ab}\pa_a\left[ \int_{\eta'}^{\hat\eta_o}d\eta''\Ccal_{rr} \right]
\pa_b\left[ \int_{\eta'}^{\hat\eta_o}d\eta''\Ccal_{rr} \right]
-2\,\bar\ga^{ab}\,\Ccal_{ar} \pa_b\int_{\eta'}^{\hat\eta_o}d\eta''\Ccal_{rr}\right\}\nonumber\\
&
+\left[\delta x^\mu_z\frac{\pa\delta Z^{(1)}(\hat x^\mu_o,\hat x^\mu_z)}{\pa \hat x^\mu_z}\right]_{SG}\,,
\label{eq:dz_SG0}
\end{align}
or, after combining with the decompositions in Eqs. \eqref{eq:SG_dec}
\begin{align}
&\delta Z^{(2)}_{SG}(\hat x^\mu_o,\hat x^\mu_z)=\,
-\int_{\hat\eta_z}^{\hat\eta_o}d\eta'\,\pa_{\eta'}\left\{  -\psi
-\pa^2_r\ga
-\pa_rC^{(2)}_r
-C^{(2)}_{rr}
+\frac{1}{2}\,\left( \psi+\pa^2_r\ga \right)^2\right.\nonumber\\
&+2\,\bar\ga^{ab}\left(\pa^2_{ra}\ga-\frac{1}{r}\pa_a\ga\right)\,\left(\pa^2_{rb}\ga-\frac{1}{r}\pa_b\ga\right)
-\left( \psi+\pa^2_r\ga \right)\int_{\eta'}^{\hat\eta_o}d\eta''\,\pa_{\eta''}\left( \psi+\pa^2_r\ga \right)
\nonumber\\
&+\frac{1}{2}\bar\ga^{ab}\pa_a\left[\int_{\eta'}^{\hat\eta_o}d\eta''\left(\psi
+\pa^2_r\ga\right)\right]\,\pa_b\left[\int_{\eta'}^{\hat\eta_o}d\eta''\left(\psi
+\pa^2_r\ga\right)\right]\nonumber\\
&\left.-2\,\bar\ga^{ab}\left(\pa^2_{ra}\ga-\frac{1}{r}\pa_a\ga\right)\pa_b\left[\int_{\eta'}^{\hat\eta_o}d\eta''\left(\psi
+\pa^2_r\ga\right)\right]\right\}
+\left[\delta x^\mu_z\frac{\pa\delta Z^{(1)}(\hat x^\mu_o,\hat x^\mu_z)}{\pa \hat x^\mu_z}\right]_{SG}\,,
\label{eq:dz_SG1}
\end{align}
where
\begin{align}
\delta t_{z\,SG}&=-\frac{1}{H}\int_{\hat\eta_z}^{\hat\eta_o}d\eta'\,\pa_{\eta'}\left(\psi+\pa^2_r\ga\right)+O(2)\nonumber\\
\delta r_{z\,SG}&=-\int_{\hat\eta_z}^{\hat \eta_o}\,d\eta\,\left( \psi+\pa^2_r\ga \right)-\frac{\delta t_{z\,SG}}{a(\hat t_z)}+O(2)\nonumber\\
\delta\theta^a_{z\,SG}&=\int_{\hat\eta_z}^{\hat \eta_o}d\eta\left[ \bar\ga^{ab}\pa_b \int_\eta^{\hat\eta_o}\,d\eta'\,\left(\psi+\pa^2_r\ga\right)
+2\,\pa^2_{rb}\ga-\frac{2}{\hat\eta_o-\eta}\,\pa_b\ga\right]+O(2)\,.
\end{align}
Term $\pa_r^2\ga$ in the first line of Eq. \eqref{eq:dz_SG1} can be integrated by part, in order to extract boundary terms which can be addressed, at linear order, to the gauge invariant part of Sachs-Wolfe and Doppler effect in SG. Indeed, using the relations
\begin{equation}
\frac{d\ga}{d\eta}=\ga'-\pa_r\ga\quad\text{and}\quad\pa_r^2\ga'=\ga'''-\frac{d}{d\eta}\left( \ga''+\pa_r\ga' \right)\,,
\end{equation}
we can re-arrange Eq. \eqref{eq:dz_SG1} as
\begin{align}
&\delta Z^{(2)}_{SG}(\hat x^\mu_o,\hat x^\mu_z)=\,
\left[\ga''
+\pa_r\ga'\right]^z_{\hat o}
-\int_{\hat\eta_z}^{\hat\eta_o}d\eta'\,\pa_{\eta'}\left\{  -\psi
-\ga''
-\pa_rC^{(2)}_r
-C^{(2)}_{rr}
+\frac{1}{2}\,\left( \psi+\pa^2_r\ga \right)^2\right.\nonumber\\
&+2\,\bar\ga^{ab}\left(\pa^2_{ra}\ga-\frac{1}{r}\pa_a\ga\right)\,\left(\pa^2_{rb}\ga-\frac{1}{r}\pa_b\ga\right)
-\left( \psi+\pa^2_r\ga \right)\int_{\eta'}^{\hat\eta_o}d\eta''\,\pa_{\eta''}\left( \psi+\pa^2_r\ga \right)
\nonumber\\
&+\frac{1}{2}\bar\ga^{ab}\pa_a\left[\int_{\eta'}^{\hat\eta_o}d\eta''\left(\psi
+\pa^2_r\ga\right)\right]\,\pa_b\left[\int_{\eta'}^{\hat\eta_o}d\eta''\left(\psi
+\pa^2_r\ga\right)\right]\nonumber\\
&\left.-2\,\bar\ga^{ab}\left(\pa^2_{ra}\ga-\frac{1}{r}\pa_a\ga\right)\pa_b\left[\int_{\eta'}^{\hat\eta_o}d\eta''\left(\psi
+\pa^2_r\ga\right)\right]\right\}
+\left[\delta x^\mu_z\frac{\pa\delta Z^{(1)}(\hat x^\mu_o,\hat x^\mu_z)}{\pa \hat x^\mu_z}\right]_{SG}\,,
\label{eq:dz_SG2}
\end{align}
which represents the full expression for non-linear redshift in the Synchronous Gauge.

Now, we can compare our results with other calculations made so far in literature. In particular, we consider the expression provided in \cite{Barausse:2005nf}. First of all, we point out two differences in solving the conformally rescaled geodesic equations: the first one is that they do not expand $\pa_\lambda k^\mu =k^\nu\pa_\nu k^\mu=\left( k^\nu_{(0)}+k^\nu_{(1)}+k^\nu_{(2)} \right)\pa_\nu\left( k^\mu_{(0)}+k^\mu_{(1)}+k^\mu_{(2)} \right)$ so their solution does not show the terms coming from $k^\nu_{(2)}\pa_\nu k^\mu_{(0)}+k^\nu_{(1)}\pa_\nu k^\mu_{(1)}$. The second difference is that they expand the liner Christoffel symbols around the linear shift $x_{(1)}^\rho\pa_\rho\Gamma^{\mu(1)}_{\alpha\beta}k^\alpha_{(0)}k^\beta_{(0)}$. This leads to some terms contains two derivatives of the perturbations which are not present in all our derivation and our gauge-transformation properties. Because of these differences, their $\delta z$ is not expected to coincide with our Eq. \eqref{eq:dz_SG2}. Moreover, we notice another conceptual difference: indeed, in \cite{Barausse:2005nf}, $\lambda$ is expanded as $\lambda=\bar\lambda+\delta\lambda$ and $\delta \lambda$ is used to relate the proper time of the source to the observed redshift, just as our $\delta t_z$ in Eq. \eqref{eq:dtG}. While the expansion around $\bar\lambda$ is legit, using it to evaluate $\delta t_z$ seems wrong, because its gauge-transformations properties do not satisfy what is required for a general coordinate time-lapse.

\subsection{Conformal Newtonian Gauge}
Now let us move the discussion of our result to the Newtonian gauge (NG), where $\Acal=\phi$ and, again, vector and tensor are considered only at the second order, so
\begin{equation}
ds^2=-(1+2\,\phi)dt^2-2\,a\,B^{(2)}_i\,dx^i\,dt+a^2\,\left[ \delta_{ij}+2\,\left(\psi\,\delta_{ij}+C^{(2)}_{ij}\right) \right]\,dx^idx^j\,,
\label{eq:NG}
\end{equation}
or, by explicitly considering the decompositions,
\begin{align}
&\qquad\qquad\Acal=\,\phi,\qquad\qquad
\Bcal_r=\,B^{(2)}_r\nonumber\\
\Ccal_{rr}=&\,\psi+C^{(2)}_{rr},\qquad\qquad
\Ccal_{ra}=\,C^{(2)}_{ra},\qquad\qquad
\Ccal_{ab}=\,\psi\,\bar\ga_{ab}+C^{(2)}_{ab}\,.
\label{eq:dec_NG}
\end{align}
First of all, this case is different from the former one because lapses are no longer null in the NG. Indeed, from Eqs. \eqref{eq:spatial_lapse} and \eqref{eq:second_time_lapses}, we have
\begin{equation}
\left(\delta x^i_{\hat o}\right)_{NG}
=-\int_{\eta_{in}}^{\hat\eta_o}d\eta\,\delta^{ij}\partial_j\int_{\eta_{in}}^\eta d\eta'\,a\,\phi\,,
\end{equation}
where we used the geodesic condition to relate $V_j$ and $\phi$, and
\begin{align}
\left(\delta t_{\hat o}\right)_{NG}=&\,-\int_{\eta_{in}}^{\hat\eta_o} d\eta\,a\,\left\{\phi-\frac{1}{2}\phi^2
+\frac{1}{2}\delta^{ij}\partial_i\left(\int_{\eta_{in}}^\eta d\eta'\,a\,\phi\right)\,\partial_j\left(\int_{\eta_{in}}^\eta d\eta''\,a\,\phi\right)\right\}
+\phi_{\hat o}\int_{\eta_{in}}^{\hat\eta_o} d\eta\,a\,\phi\nonumber\\
&+\int_{\eta_{in}}^{\hat\eta_o}d\eta\,\delta^{ij}\partial_j\left(\int_{\eta_{in}}^\eta d\eta'\,a\,\phi\right)\,\partial_i\left(\int_{\eta_{in}}^{\hat\eta_o} d\eta\,a\,\phi\right)\,.
\end{align}
In this way, the perturbed redshift is given by
\begin{align}
&\delta Z^{(2)}_{NG}(\hat x^\mu_o,\hat x^\mu_z)=\,\left\{-\Acal
+\Bcal_r+V_r
+\frac{1}{2}V_r^2
+\frac{1}{2}\bar\ga^{ab}V_aV_b
+\frac{3}{2}\,\Acal^2
-2\,V_r\,\Ccal_{rr}\right\}^z_{\hat o}
+V_{rz}\left( \Acal
-\Ccal_{rr}\right)^{\hat o}_z\nonumber\\
&-\Acal^z_{\hat o}\,\left( \Acal
-\Ccal_{rr}\right)_{\hat o}
-\left(V_r
-\Ccal_{rr} \right)_{\hat o}\left( V_r-\Acal\right)^z_{\hat o}
-\left( V_r -\Acal\right)^s_{\hat o}\int_{\hat\eta_z}^{\eta_o}d\eta'\,\pa_{\eta'}\left( \Acal
-\Ccal_{rr}\right)\nonumber\\
&-V^a_s\,\pa_a\int_{\hat\eta_z}^{\eta_o}d\eta'\,\left( \Acal
-\Ccal_{rr}\right)
-\int_{\hat\eta_z}^{\eta_o}d\eta'\,\pa_{\eta'}\left\{\Acal
-\Ccal_{rr}
-\Bcal_r
-\frac{3}{2}\Acal^2\right.
+\Acal\,\Ccal_{rr}
+\frac{1}{2}\,\Ccal^2_{rr}\nonumber\\
&-\left( \Acal-\Ccal_{rr} \right)\int_{\eta'}^{\hat\eta_o}d\eta''\,\pa_{\eta''}\left(\Acal-\Ccal_{rr}\right)
+\frac{1}{2}\bar\ga^{ab}\pa_a\left[ \int_{\eta_{in}}^{\hat\eta_o}d\eta\left( \frac{a}{a_{\hat o}}\Acal+V_r \right)-\int_{\eta'}^{\hat\eta_o}d\eta''\left( \Acal-\Ccal_{rr} \right) \right]\nonumber\\
&\left.\times\pa_b\left[ \int_{\eta_{in}}^{\hat\eta_o}d\eta\left[ \frac{a}{a_{\hat o}}\Acal+V_r \right]-\int_{\eta'}^{\hat\eta_o}d\eta''\left( \Acal-\Ccal_{rr} \right) \right]\right\}
+\left( V^a \right)^z_{\hat o}\,\pa_a \int_{\eta_{in}}^{\hat\eta_o}d\eta\left( \frac{a}{a_{\hat o}}\Acal+V_r \right)\nonumber\\
&+\delta x^\mu_{\hat o} \left(\frac{\pa\Delta\left( x^\mu_o,\hat x^\mu_z \right)}{\pa x^\mu_o}\right)_{x^\mu_o=\hat x^\mu_o}
-\delta x^\mu_{\hat o}\left(\frac{\pa_\mu\bar\Ecal}{\bar\Ecal}\right)_{\hat o}
-\frac{1}{2}\delta x^\mu_{\hat o}\delta x^\nu_{\hat o}\left(\frac{\pa^2_{\mu\nu}\bar\Ecal}{\bar\Ecal}\right)_{\hat o}+\left( \delta x^\mu\frac{\pa_\mu\bar\Ecal}{\bar\Ecal} \right)_{\hat o}^2\nonumber\\
&-\Delta\left( \hat x^\mu_o,\hat x^\mu_z \right)\delta x^\mu_{\hat o}\left(\frac{\pa_\mu\bar\Ecal}{\bar\Ecal}\right)_{\hat o}
+\left[\delta x^\mu_z\frac{\pa\delta Z^{(1)}(\hat x^\mu_o,\hat x^\mu_z)}{\pa \hat x^\mu_z}\right]_{NG}\,,
\label{eq:dz_second_NG}
\end{align}
where
\begin{align}
\delta t_{z\,NG}=&\,\frac{1}{H}\left[ \left(-\phi+V_r \right)^z_{\hat o}
-\int_{\hat\eta_z}^{\hat\eta_o}d\eta'\,\pa_{\eta'}\left( \phi
-\psi \right)
-\frac{\Hcal_{\hat o}}{a_{\hat o}}\int_{\eta_{in}}^{\hat\eta_o}d\eta\,a\,\phi \right]+O(2)\nonumber\\
\delta r_{z\,NG}&=\int_{\hat\eta_z}^{\hat \eta_o}\,d\eta\,\left( \phi-\psi \right)+\frac{\left(\delta t_{\hat o}\right)_{NG}}{a(\hat t_o)}-\frac{\delta t_{z\,NG}}{a(\hat t_z)}+\left(\delta r_{\hat o}\right)_{NG}+O(2)\nonumber\\
\delta\theta^a_{z\,NG}&=-\int_{\hat\eta_z}^{\hat \eta_o}d\eta\left\{ \bar\ga^{ab}\pa_b\left[ \int_\eta^{\hat\eta_o}\,d\eta'\,\left( \phi-\psi \right)+\frac{\left(\delta t_{\hat o}\right)_{NG}}{a_{\hat o}}+\left(\delta r_{\hat o}\right)_{NG} \right]\right\}+\left(\delta\theta^a_{\hat o}\right)_{NG}+O(2)\,.
\end{align}
By inserting Eqs. \eqref{eq:dec_NG} into Eq. \eqref{eq:dz_second_NG}, we get that linear redshift is then given by
\begin{equation}
\delta Z^{(1)}_{NG}(\hat x^\mu_o,\hat x^\mu_z)=\,\left(-\phi+V_r \right)^z_{\hat o}
-\int_{\hat\eta_z}^{\hat\eta_o}d\eta'\,\pa_{\eta'}\left( \phi
-\psi \right)
-\frac{\Hcal_{\hat o}}{a_{\hat o}}\int_{\eta_{in}}^{\hat\eta_o}d\eta\,a\,\phi\,,
\label{eq:NG_linear}
\end{equation}
and the second-order is
\begin{align}
&\delta Z^{(2)}_{NG}(\hat x^\mu_o,\hat x^\mu_z)
=\,\left[-\phi
+B^{(2)}_r
+V_r
+\frac{1}{2}\bar\ga^{ab}V_a\,V_b\right]^z_{\hat o}
+\frac{1}{2}\left( V_{r\,z}-V_{r\,\hat o} \right)^2\nonumber\\
&+\left(\phi-V_r\right)^z_{\hat o}\int_{\hat \eta_z}^{\hat\eta_o}d\eta'\pa_{\eta'}\left(\phi-\psi\right)
-V^a_z\,\pa_a\int_{\hat\eta_z}^{\hat\eta_o}d\eta'\left( \phi-\psi \right)
+\frac{3}{2}\phi^2_z-\frac{1}{2}\phi^2_{\hat o}
-\phi_z\,\phi_{\hat o}
-\left(\phi\right)^z_{\hat o}\left(V_r\right)^z_{\hat o}\nonumber\\
&-\left( V_r\,\psi \right)^z_{\hat o}
-\int_{\hat\eta_z}^{\hat\eta_o}d\eta\,\pa_\eta\left[\phi-\psi
-C^{(2)}_{rr}
-B^{(2)}_r
-\frac{3}{2}\phi^2
+\frac{1}{2}\,\psi^2
+\phi\,\psi
-\left(\phi-\psi\right)\int_\eta^{\hat\eta_o}d\eta'\,\pa_{\eta'}\left( \phi-\psi \right)\right.\nonumber\\
&\left.+\frac{1}{2}\bar\ga^{ab}\pa_a\int_\eta^{\eta_o}d\eta'\left( \phi-\psi \right)\,\pa_b\int_\eta^{\hat\eta_o}d\eta''\left( \phi-\psi \right)\right]\nonumber\\
&-\frac{1}{2}\int_{\hat\eta_z}^{\hat\eta_o}d\eta\,\partial_\eta\left\{\bar\ga^{ab}\pa_a\int_{\eta_{in}}^{\hat\eta_o}d\eta\left( \frac{a}{a_{\hat o}}\phi+V_r \right)\,\pa_b\int_{\eta_{in}}^{\hat\eta_o}d\eta\left( \frac{a}{a_{\hat o}}\phi+V_r \right)\right.\nonumber\\
&\left.-\bar\ga^{ab}\pa_a\int_{\eta_{in}}^{\hat\eta_o}d\eta\left( \frac{a}{a_{\hat o}}\phi+V_r \right)\,\pa_b\int_\eta^{\hat\eta_o}d\eta'\left( \phi-\psi \right)\right\}
+\left( V^a \right)^z_{\hat o}\,\pa_a \int_{\eta_{in}}^{\hat\eta_o}d\eta\left( \frac{a}{a_{\hat o}}\phi+V_r \right)\nonumber\\
&+\delta x^\mu_{\hat o} \left(\frac{\pa\Delta\left( x^\mu_o,\hat x^\mu_z \right)}{\pa x^\mu_o}\right)_{x^\mu_o=\hat x^\mu_o}
-\delta x^\mu_{\hat o}\left(\frac{\pa_\mu\bar\Ecal}{\bar\Ecal}\right)_{\hat o}
-\frac{1}{2}\delta x^\mu_{\hat o}\delta x^\nu_{\hat o}\left(\frac{\pa^2_{\mu\nu}\bar\Ecal}{\bar\Ecal}\right)_{\hat o}+\left( \delta x^\mu\frac{\pa_\mu\bar\Ecal}{\bar\Ecal} \right)_{\hat o}^2\nonumber\\
&-\Delta\left( \hat x^\mu_o,\hat x^\mu_z \right)\delta x^\mu_{\hat o}\left(\frac{\pa_\mu\bar\Ecal}{\bar\Ecal}\right)_{\hat o}
+\left[\delta x^\mu_z\frac{\pa\delta Z^{(1)}(\hat x^\mu_o,\hat x^\mu_z)}{\pa \hat x^\mu_z}\right]_{NG}\,.
\end{align}
In order to facilitate the comparison with literature about second order \cite{Marozzi:2014kua,Umeh:2014ana,Pyne:1995bs}, let us assume vanishing anisotropic stress, i.e. $\psi=-\phi$ at linear order and in the quadratic terms. We then get
\begin{align}
&\delta Z^{(2)}_{NG}(\hat x^\mu_o,\hat x^\mu_z)
=\,\left[-\phi
+B^{(2)}_r
+V_r
+\frac{1}{2}\bar\ga^{ab}V_a\,V_b\right]^z_o
+\frac{1}{2}\left( V_{r\,z}-V_{r\,o} \right)^2\nonumber\\
&+2\,\left(\phi-V_r\right)^z_o\int_{\hat\eta_z}^{\hat\eta_o}d\eta'\phi'
-2\,V^a_z\,\pa_a\int_{\hat\eta_z}^{\hat\eta_o}d\eta'\phi
+\frac{3}{2}\phi^2_z-\frac{1}{2}\phi^2_o
-\phi_z\,\phi_o
-\left(\phi\right)^z_o\left(V_r\right)^z_o+\left( V_r\,\phi \right)^z_o\nonumber\\
&-\int_{\hat\eta_z}^{\hat\eta_o}d\eta\,\pa_\eta\left[\phi-\psi
-C^{(2)}_{rr}
-B^{(2)}_r
-2\,\phi^2
-4\,\phi\int_\eta^{\hat\eta_o}d\eta'\phi'
+2\,\bar\ga^{ab}\pa_a\left(\int_\eta^{\hat\eta_o}d\eta'\phi\right)\,\pa_b\left(\int_\eta^{\hat\eta_o}d\eta''\phi\right) \right]\nonumber\\
&-\frac{1}{2}\int_{\hat\eta_z}^{\hat\eta_o}d\eta\,\partial_\eta\left\{\bar\ga^{ab}\pa_a\int_{\eta_{in}}^{\hat\eta_o}d\eta\left( \frac{a}{a_{\hat o}}\phi+V_r \right)\,\pa_b\int_{\eta_{in}}^{\hat\eta_o}d\eta\left( \frac{a}{a_{\hat o}}\phi+V_r \right)\right.\nonumber\\
&\left.-2\,\bar\ga^{ab}\pa_a\int_{\eta_{in}}^{\hat\eta_o}d\eta\left( \frac{a}{a_{\hat o}}\phi+V_r \right)\,\pa_b\int_\eta^{\hat\eta_o}d\eta' \phi\right\}
+\left( V^a \right)^z_{\hat o}\,\pa_a \int_{\eta_{in}}^{\hat\eta_o}d\eta\left( \frac{a}{a_{\hat o}}\phi+V_r \right)\nonumber\\
&+\delta x^\mu_{\hat o} \left(\frac{\pa\Delta\left( x^\mu_o,\hat x^\mu_z \right)}{\pa x^\mu_o}\right)_{x^\mu_o=\hat x^\mu_o}
-\delta x^\mu_{\hat o}\left(\frac{\pa_\mu\bar\Ecal}{\bar\Ecal}\right)_{\hat o}
-\frac{1}{2}\delta x^\mu_{\hat o}\delta x^\nu_{\hat o}\left(\frac{\pa^2_{\mu\nu}\bar\Ecal}{\bar\Ecal}\right)_{\hat o}+\left( \delta x^\mu\frac{\pa_\mu\bar\Ecal}{\bar\Ecal} \right)_{\hat o}^2\nonumber\\
&-\Delta\left( \hat x^\mu_o,\hat x^\mu_z \right)\delta x^\mu_{\hat o}\left(\frac{\pa_\mu\bar\Ecal}{\bar\Ecal}\right)_{\hat o}
+\left[\delta x^\mu_z\frac{\pa\delta Z^{(1)}(\hat x^\mu_o,\hat x^\mu_z)}{\pa \hat x^\mu_z}\right]_{NG}\,.
\end{align}
The second order part can be split into several terms as already done in \cite{Marozzi:2014kua,Umeh:2014ana}. In particular, we split
\begin{align}
\delta Z^{(2)}_{NG}\equiv&\,\delta Z_S+\delta Z_{Doppler}+\delta Z_{SW}+\delta Z_{IISW}+\delta Z_{SW\times ISW}\nonumber\\
&+\delta Z_{Doppler\times ISW}+\delta Z_{Doppler\times SW}+\delta Z_{\hat o}+\delta Z_{z}\,,
\end{align}
where we define
\begin{align}
\delta Z_{S}\equiv&\left[-\phi^{(2)}+B^{(2)}_r+V^{(2)}_r \right]^z_o-\int_{\hat\eta_z}^{\hat\eta_o}d\eta\,\left(\phi^{(2)}-\psi^{(2)}
-C^{(2)}_{rr}
-B^{(2)}_r\right)'\nonumber\\
\delta Z_{Doppler}\equiv&\left[\frac{1}{2}\bar\ga^{ab}V_a\,V_b\right]^z_o
+\frac{1}{2}\left(V_{r\,z}-V_{r\,o}\right)^2\nonumber\\
\delta Z_{SW}\equiv&\,\frac{3}{2}\phi^2_z-\frac{1}{2}\phi^2_o
-\phi_z\,\phi_o\nonumber\\
\delta Z_{IISW}\equiv&-\int_{\hat\eta_z}^{\hat\eta_o}d\eta\,\pa_\eta\left[
-2\,\phi^2
-4\,\phi\int_\eta^{\hat\eta_o}d\eta'\phi'
+2\,\bar\ga^{ab}\pa_a\left(\int_\eta^{\hat\eta_o}d\eta'\phi\right)\,\pa_b\left(\int_\eta^{\hat\eta_o}d\eta''\phi\right) \right]\nonumber\\
\delta Z_{SW\times ISW}\equiv&\,2\,\left(\phi_z-\phi_o \right)\int_{\hat\eta_z}^{\hat\eta_o}d\eta'\phi'\nonumber\\
\delta Z_{Doppler\times ISW}\equiv&\,-2\,\left( V_{r\,z}-V_{r\,o} \right)\int_{\hat\eta_z}^{\hat\eta_o}d\eta'\phi'
-2\,V^a_z\,\pa_a\int_{\hat\eta_z}^{\hat\eta_o}d\eta'\phi\nonumber\\
\delta Z_{Doppler\times SW}\equiv&\,-\left(\phi\right)^z_o\left(V_r\right)^z_o+\left( V_r\,\phi \right)^z_o
=V_{r\,o}\left( \phi_z-\phi_o \right)+\phi_o\left( V_{r\,z}-V_{r\,o} \right)\nonumber\\
\delta Z_{\hat o}\equiv
&\,-\frac{1}{2}\int_{\hat\eta_z}^{\hat\eta_o}d\eta\,\partial_\eta\left\{\bar\ga^{ab}\pa_a\int_{\eta_{in}}^{\hat\eta_o}d\eta\left( \frac{a}{a_{\hat o}}\phi+V_r \right)\,\pa_b\int_{\eta_{in}}^{\hat\eta_o}d\eta\left( \frac{a}{a_{\hat o}}\phi+V_r \right)\right.\nonumber\\
&\left.-2\,\bar\ga^{ab}\pa_a\int_{\eta_{in}}^{\hat\eta_o}d\eta\left( \frac{a}{a_{\hat o}}\phi+V_r \right)\,\pa_b\int_\eta^{\hat\eta_o}d\eta' \phi\right\}\nonumber\\
&+\left( V^a \right)^z_{\hat o}\,\pa_a \int_{\eta_{in}}^{\hat\eta_o}d\eta\left( \frac{a}{a_{\hat o}}\phi+V_r \right)
+\delta x^\mu_{\hat o} \left(\frac{\pa\Delta\left( x^\mu_o,\hat x^\mu_z \right)}{\pa x^\mu_o}\right)_{x^\mu_o=\hat x^\mu_o}
-\delta x^\mu_{\hat o}\left(\frac{\pa_\mu\bar\Ecal}{\bar\Ecal}\right)_{\hat o}\nonumber\\
&-\frac{1}{2}\delta x^\mu_{\hat o}\delta x^\nu_{\hat o}\left(\frac{\pa^2_{\mu\nu}\bar\Ecal}{\bar\Ecal}\right)_{\hat o}+\left( \delta x^\mu\frac{\pa_\mu\bar\Ecal}{\bar\Ecal} \right)_{\hat o}^2-\Delta\left( \hat x^\mu_o,\hat x^\mu_z \right)\delta x^\mu_{\hat o}\left(\frac{\pa_\mu\bar\Ecal}{\bar\Ecal}\right)_{\hat o}\nonumber\\
\delta Z_z\equiv&\,\left[\delta x^\mu_z\frac{\pa\delta Z^{(1)}(\hat x^\mu_o,\hat x^\mu_z)}{\pa \hat x^\mu_z}\right]_{NG}
\end{align}
Then $\delta Z^{(2)}_{NG}$ is given by the sum of all the previous terms and $\delta z_S$ is the redshift entirely sourced by pure second order perturbations, $\delta Z_{Doppler}$ is the redshift sourced by quadratic linear order peculiar velocity, $\delta Z_{SW}$ is the redshift perturbation due to local quadratic potentials, $\delta Z_{IISW}$ is the second order and double integrated Sachs-Wolfe effect and all the others are just cross terms among all the previous effect. Let us also notice the presence of an extra term $\delta Z_{\hat o}$, due to the coordinate lapse and shift $\delta t_o$ and $\delta x^i_o$ and normalization constant $\delta w_o$, which are not present in the other papers. Moreover, we have also the terms $\delta Z_z$ which involves coordinate lapse and shift at the source because we express everything in terms of observed redshift, past light-cone and angles.

Now, we have all the ingredients to compare our results with the previous ones in literature\footnote{In this comparison, we refer to formula as they are enumerated in the final arXiv versions}. The very first attempt of evaluating redshift perturbations in literature has been performed in \cite{Pyne:1995bs}. From a direct comparison at linear order between our Eq. \eqref{eq:NG_linear} and their Eq. (4.11), we can infer that their unspecified function $\tau$ corresponds to the peculiar velocity $V_r$. Moreover, at second order, we notice that authors in \cite{Pyne:1995bs} have been interested in considering only non linear terms from quadratic perturbations (see their Eq. (4.12)). This fact, combined with their neglecting of the time lapse and spatial shift at the observer, allows us to provide only a very partial comparison of the results.

A more detailed comparison can be provided with the result obtained in the GLC literature \cite{BenDayan:2012wi,Marozzi:2014kua,Fanizza:2013doa,BenDayan:2013gc,Fanizza:2015swa}. Whereas \cite{BenDayan:2012wi,Fanizza:2013doa,BenDayan:2013gc,Fanizza:2015swa} provide only the expression for $\delta\Ecal$ (or, equivalently $\delta \U$), which is the key quantity for evaluating the redshift, a direct comparison can be made with Eqs. (4-11)-(4.20) of \cite{Marozzi:2014kua}, where $\delta Z$ has been explicitly written down. Beside the absence of $\delta Z_{\hat o}$ and the terms proportional to $\delta \tilde\theta^a_{\hat o}$ and $\delta w_o$ in the angular part of $\delta Z_z$, we notice a good agreement with almost all the terms. The only evident disagreement is about the sub-leading terms $\left( \phi\,V_r \right)^z_{\hat o}$ in $\delta Z_{Doppler\times SW}$, which is present only in our derivation. However, we notice that our expression for $\delta\Upsilon$ agrees with all the ones derived in GLC literature, which seems to agree with our expression of $\delta Z_{Doppler\times SW}$. Moreover, the expansion around the reference frame at background agrees only partially with \cite{Marozzi:2014kua}: indeed, $\delta Z_z$ can be expanded as
\begin{equation}
\delta Z_z=\left(\frac{\delta Z^{(1)}}{H}\dot{\delta Z^{(1)}}+\delta r_z\pa_r\delta Z^{(1)}+\delta \theta^a_z\pa_a\delta Z^{(1)}\right)_{NG}\,,
\end{equation}
so only the final expansion around the observed angles is present there (namely their $\delta z_{\delta\theta}$), which contains the leading contributions. However, the full expansion around $\hat t_z$ and $\hat r_z$ too is always considered in all the GLC paper in the final step where relations between physical observables (for instance $d_L(z)$) are considered. Finally, our term for $\delta Z_{IISW}$ agrees with the one of \cite{Marozzi:2014kua} as presented in its Eq. (3.9).

Last comparison we can perform is with \cite{Umeh:2014ana}. The authors here consider all the possible corrections from observer and source positions. Moreover, they also take into account perturbations arising from induced vectors and tensors at second order. Also here, we found a disagreement in $\delta Z_{Doppler\times SW}$. Indeed, their Eq. (83) can be recombined and give a term which, in our notation, is $\left( \phi\,V_r \right)^z_{\hat o}+\left( \phi_z+\phi_{\hat o} \right)\left( V_r \right)^z_{\hat o}$. Hence, the term $\phi_z\left( V_r \right)^z_{\hat o}$ has opposite sign with respect to our results. On the other hand, also $\delta z_{IISW}$ seems to differ. We notice that \cite{Umeh:2014ana} does not take into account both $\delta Z_{\hat o}$ and any of the $\delta Z_z$ terms while we agree for perturbations due to induced vectors and tensors.

\subsection{Summary of the comparison}
We conclude this section by providing a summary of our comparison 
with the other second-order calculations of the observed redshift.
First of all, the observer terms due to the time lapse and spatial shift 
have never been considered in literature, except our recent work.
In addition, the main differences can be summarized as follows:
\begin{itemize}
\item{Baruasse et al.: we point out three difference in the procedure of solving the photon geodesic equation. First of all the authors solve the geodesic equation in terms of an affine parameter which is not expanded around the perturbed photon's path from the source to the observer. This implies that their expression misses all the terms coming from the expansion of the perturbed geodesic around the unperturbed one, i.e. the post-Born effects. Secondly they try to recover these terms by expanding the Christoffel symbols around the unperturbed geodesic. This procedure generates some terms which couples to the derivatives of the $\Gamma$ which never appear in our procedure. The last difference is in the definition of the source time shift which is needed to express other physical observables in term of the observed redshift. They expand the affine parameter around the unperturbed geodesic to build it. However this shift does not satisfy the required transformation property of a generic time shift exploited in Sect. \ref{sec:GI} which is needed to get the gauge invariance for a given observables in term of the redshift.}
\item{Pyne \& Carrol: the derivation presented in that work is very partial and does not allow a full comparison with our result beyond the linear order. Indeed they consider only quadratic perturbations in the expression for the redshift.}
\item{Marozzi: we underline a difference in the term $\delta Z_{Doppler\times SW}$. Indeed our expression has an extra term $(\phi V_r)^z_o$ which combines with $(\phi)^z_o\,(V_r)^z_o$ and provide a cancellation of the source term $\phi_z\,V_{r\,z}$. Moreover we underline that in Marozzi the full expansion around the observed redshift is made directly for the final relation $d_L(z)$ and not for the redshift itself where only the expansion in terms of the observed angles is provided. The remaining terms for the redshift are in agreement with our derivation.}
\item{Umeh et al.: here we find full agreement for the perturbations induced by vector and tensor up to second order. However, we underline a disagreement with our derivation in the term $\delta Z_{Doppler\times SW}$: indeed they found $(\phi\,V_r)^z_o+(\phi_z+\phi_o)(V_r)^z_o$ while our result is $(\phi\,V_r)^z_o+(-\phi_z+\phi_o)(V_r)^z_o$. Even in this case, this difference leads to a combination which does not cancel the source contribution $\phi_z\,V_{r\,z}$. Moreover we notice that this result does not take into account at all the expansion around the observed angles. Finally the term $\delta Z_{IISW}$ looks different from our result.}
\end{itemize}

\section{Summary and Discussion}
\label{sec:conclusion}
In this paper, we have provided the second-order expression of the
observed redshift with all the relativistic effects taken into consideration
and verified the gauge-transformation property of its expression.
Our result marks the first explicit calculations of otherwise difficult
second-order relativistic perturbations in deriving its expression
without choosing a gauge condition and in checking its gauge transformation.
The observable quantities should follow a unique transformation law,
and by explicit calculations of gauge-transformation we have verified
that our second-order expressions indeed satisfy this condition,
and this consistency check provides the strongest support ever that our
calculations are correct. Furthermore, since our expression is general,
we can freely impose any gauge conditions, facilitating easy comparison
of our result to previous work in literature, and
we have provided detailed comparisons in several popular
choices of gauge conditions.

While checking the gauge invariance is a powerful way to ensure the sanity
of relativistic perturbation calculations, there exists a
subtlety associated with the observable quantities.
The observable quantities are expressed in terms of diffeomorphism
invariant scalars, i.e., invariant under a coordinate transformation.
However, they still gauge transform, because they are evaluated at
the same coordinate value in two different coordinate systems;
in other words,
these observables are evaluated at two different physical positions.
In fact, the physical points~$P$ of our interest such as the observer
position or the source position are expressed in terms of
coordinate-independent reference $\hat x^\mu_o = (\bar t_o,0)$
for the observer position and $\hat x^\mu_s=(\hat t_z,\hat r_z n^i)$
for the source position, where $\bar t_o$ is related to the
proper time of the observer in Eq.~\eqref{eq:35} and $\hat t_z$ and~$\hat r_z$ to the
observed redshift in Eqs.~\eqref{eq:419}. These reference points~$\hat x^\mu_P$
are important
in expressing the physical points of our interest in a coordinate
independent way.

However, though these reference points~$\hat x^\mu_P$
have the same values in any coordinate
systems, their position in each coordinate system represents points
different from the physical points they are designed to refer to. Instead,
these physical points~$P$ in each coordinate differ from the reference
points by a coordinate lapse and shift: $x^\mu_P=\hat x^\mu_P+\delta x^\mu_P$,
each of which describes the difference in the spacetime coordinate values
between the physical and the reference points in a coordinate system.
By splinting the position into the (coordinate-independent) background
and the perturbation, we shift the coordinate transformation of the physical
points to the coordinate lapse and shift, and the background reference
remains unaffected,
as we do in the perturbation analysis. Therefore, by evaluating the observable
quantities at the reference points and adding the coordinate lapse and
shift, we practically evaluate the observable quantities at the same physical
points in any coordinate systems, systematically achieving the goal of
diffeomorphism invariance. Indeed, this approach has been adopted, albeit
not explicitly stated or understood,
in most of linear-order perturbation analysis. However, beyond linear order,
the role of the reference points and the coordinate lapse and shift is
critical, and we clarified how we can evaluate the perturbation expressions
and check their gauge-invariance by expanding around the reference points.

Since the background quantities depend only on time, the coordinate shift
drops out in the linear-order expressions of the observable quantities,
and only the coordinate lapse plays a physical role.
In particular, the coordinate lapse at the source position is frequently
computed, again though not explicitly stated, at the linear-order
calculations of the observable quantities such as the luminosity distance,
and so on. At the same footing, however,
the coordinate lapse at the observer
position is as important at the linear order, and its absence
breaks the gauge-invariance and causes the infrared divergences in
numerical calculations (see, e.g., \cite{Biern:2016kys}). Beyond the linear order,
we showed that
both the coordinate lapse and the spatial shift are important in deriving
the gauge-invariant expressions of the observable quantities.
Our application of this systematic approach to higher-order perturbation
analysis in this paper was to derive the gauge-invariant expression for
the observed redshift at the second order, as it is the simplest
among other second-order calculations of the observable quantities.
Having clarified the procedure in a systematic way, we will proceed
to derive the luminosity distance, galaxy clustering, and other observable
quantities.

Hence in this work we have discussed in details the significance of these terms in order to ensure the correct gauge transformation properties, not only for the redshift itself, but also for a generic physical observable when expressed in terms of the observed redshift. 
The question then naturally arises: what are the impact of the newly found
perturbation contributions on cosmological observables. The answer to this
question has {\it no} single answer, because it depends on what
we measure. For instance, consider
the perturbation contributions at the observer position, which are often
neglected in the previous calculations (note that these are not the only 
ones). These terms at the observer position only contribute to the low
angular multipoles of $l\leq2$, as the tensor perturbations at the observer
position can contract with two line-of-sight directions to form a scalar.
Therefore, in the absence of primordial gravitational waves, 
these observer terms have {\it no impact} on the angular power spectrum.
In a dramatic contrast, it was already shown in \cite{Biern:2016kys}
that these terms are absolutely needed to cancel the infrared divergence
in computing the variance of the luminosity distance. In other words,
the correct (dimensionless) variance in the luminosity distance is
a percent level and remains constant beyond the horizon scale, while the previous calculations without the terms at the
observer position need the introduction of an infrared cutoff on the horizon scale for the variance in order to get a finite result. Hence the error made in the previous calculations is {\it cutoff dependent}. If we consider
the two-point correlation function \cite{Scaccabarozzi:2018vux}, the observer
terms give rise to {\it a few percent} corrections to 
the galaxy two-point correlation function beyond the baryonic acoustic
oscillation scale (however, note that the potential contributions to
the two-point correlation function in previous calculations are again 
{\it cutoff dependent} and hence the error made there involves the choice of the cutoff).

Furthermore, we need to emphasize that the above discussion is rather limited
to the linear-order calculations and 
the case for our second-order calculations are indeed more complicated.
As we demonstrated in our paper, we need the perturbation terms at the
observer position (and others as well) to ensure that the full expression
is gauge invariant and to prevent any infrared divergences. Moreover,
in the second-order calculations,
those observer terms now couple to the perturbation contributions at
the source position or along the line-of-sight direction, such that
they {\it do} contribute to all angular multipoles, i.e., they do not drop
out in higher angular multipoles, as they do in the linear-order calculations.
For example, consider the effect of the observer motion on the CMB multipoles.
It was shown \cite{Challinor:2002zh} that this effect
provides a correction of about 0.1\%. Given that the upcoming surveys
aim to achieve precision measurements of various cosmological observables
at the sub-percent level, it is important that the theoretical calculations
meet the strict demand set by these upcoming surveys. However,
it requires extensive investigations
of various observables to quantify the impact of the perturbation 
contributions neglected in the previous calculations, which is
beyond our current scope.

To conclude, we want to underline that the last term in Eq. \eqref{eq:dz_second} contains also the expansion around the observed angles. In literature \cite{Marozzi:2014kua,Umeh:2014ana} this term is already included in the definition of $\delta Z$. The expression of $\delta\theta^a_z$ as given in Eq. \eqref{eq:dxG} transforms under a gauge transformation as Eq. \eqref{eq:source_time_shift} thanks to the careful normalization of the observer terms, as pointed out in \cite{Scaccabarozzi:2017ncm} and more carefully analyzed in Appendix \ref{app:GLC}. Even though this term is a pure non-linear effect, it can be important in some relevant cases. For instance, let us consider the two point correlation function for temperature anisotropies in the CMB: it is well-known that the redshift perturbations are related the fluctuation $\Delta T/T$ in temperature. On the other hand, because the term $\delta x^\mu_z\pa_\mu \delta Z^{(1)}$ contains the expansion around the observed angles $\delta \theta^a_z\pa_a\delta Z^{(1)}$, this is nothing but $\delta \theta^a_z\pa_a\Delta T/T$. This correction leads to the lensed CMB spectrum for temperature anisotropies and it affects the unlensed spectrum of the two point correlation function of $\Delta T/T$ about $10\%$ on small scales ($\ell>1000$) \cite{Lewis:2006fu,RuthBook,Marozzi:2016uob,Lewis:2016tuj}. In this paper, we provided the correct second-order expression for the observed redshift and showed that there exist many terms at the source and the observer positions that are not properly accounted for. Their impact on the observed CMB power spectra is left for a future work.

\section*{Acknowledgements}
We thank Ruth Durrer, Giovanni Marozzi and Fulvio Scaccabarozzi for useful discussions and Julian Adamek for discussion on the N-body gauge properties. GF and JY acknowledge support by the Swiss National
Science Foundation and by a Consolidator Grant of the European Research Council (ERC-2015-CoG grant 680886).

\appendix
\section*{Appendix}
\section{Notation convention and Geodesic Light-Cone gauge}
\subsection{FRW metric convention}
\label{app:metric}
We consider a flat Friedmann-Robertson-Walker (FRW) metric with small perturbations in cartesian coordinates
\begin{equation}
ds^2=-(1+2\,\mathcal{A})dt^2-2\,a\,\mathcal{B}_i\,dx^i\,dt+a^2\,\left( \delta_{ij}+2\,\mathcal{C}_{ij} \right)\,dx^idx^j\,,
\label{eq:gauge_ready_app}
\end{equation}
where the perturbation variables are meant to represent the deviation from the background up to second order, e.g., $\Acal=\Acal^{(1)}+\Acal^{(2)}$. According to the transformation properties, we further decompose all the perturbations as
\begin{align}
\mathcal{A}=\,\phi\,,\qquad\qquad
\mathcal{B}_i=\,\pa_i\beta+B_i\qquad\text{and}\qquad
\mathcal{C}_{ij}=\,\psi\,\delta_{ij}+\pa^2_{ij}\gamma+\pa_{(i}C_{j)}+C_{ij}\,,
\label{eq:decompositions}
\end{align}
where $\phi,\,\beta,\,\psi$ and $\gamma$ are scalars, $B_i$ and $C_i$ are transverse vectors and $C_{ij}$ is a symmetric, transverse and traceless tensor\footnote{Here we denote scalar, vector and tensor with respect to the background symmetry $SO(3)$ in spatial coordinates.}. We also express the FRW metric in spherical coordinates to facilitate comparison to the calculations in Geodesic Light-Cone (GLC) gauge, so that the line element in this case becomes
\begin{align}
ds^2=&\,-(1+2\,\mathcal{A})dt^2-2\,a\,\mathcal{B}_r\,dr\,dt-2\,a\,\mathcal{B}_a\,dy^a\,dt\nonumber\\
&+a^2\,\left( 1+2\,\mathcal{C}_{rr} \right)\,dr^2
+2\,a^2\,\mathcal{C}_{ra}\,drdy^a
+a^2\,\left( \bar\ga_{ab}+2\,\mathcal{C}_{ab} \right)\,dy^ady^b\,,
\label{eq:gauge_ready_polar}
\end{align}
where $y^a=\left( \theta,\phi \right)$ is an angular coordinate and $\bar\ga_{ab}=r^2\,\text{diag}\left( 1,\sin^2\theta \right)$ is the metric in a 2D sphere with radius $r$. The metric perturbations are again decomposed in terms of scalar, vector and tensor as
\begin{align}
\Bcal_r=&\,\pa_r\beta+B_r\,,\qquad\qquad
\Bcal_a=\,\pa_a\beta+B_a\,,\nonumber\\
\Ccal_{rr}=&\,\psi+\pa^2_r\ga+\pa_rC_r+C_{rr}\,,\nonumber\\
\Ccal_{ra}=&\,\pa^2_{ra}\ga-\frac{1}{r}\pa_a\ga+\pa_{(a}C_{r)}-\frac{1}{r}C_a+C_{ra}\,,\nonumber\\
\Ccal_{ab}=&\,\psi\,\bar\ga_{ab}+\nabla_a\pa_b\ga+\nabla_{(a}C_{b)}+C_{ab}\,.
\end{align}

\subsection{Geodesic Light Cone coordinates and the second-order expressions}
\label{app:GLC}
Here we consider the so called Geodesic Light Cone (GLC) gauge, firstly introduced in \cite{Gasperini:2011us}, in which the line element looks like:
\begin{equation}
ds^2=\Upsilon^2dw^2-2\,\Upsilon dw\,d\tau+\gamma_{ab}\left( d\tilde\theta^a-U^adw \right)\left( d\tilde\theta^b-U^bdw \right)\quad,\quad a,b=1,2
\label{eq:GLC}
\end{equation}
where $\gamma_{ab}$ is a symmetric $2\times2$ tensor. This line element is exact in the sense that no symmetries are imposed in it. The only constraints regard the gauge fixing allowed by diffeomorphism invariance in general relativity. This is evident from the metric in which we have six arbitrary functions depending on all the four coordinates(see \cite{Fanizza:2013doa,Fleury:2016htl} for the construction and the discussion about the geometrical properties of this line element). The physical advantage of the line element in Eq. \eqref{eq:GLC} is that the photon geodesic equations can be exactly solved in the GLC gauge: the exact quadri-momentum for photon is $k_\mu=\pa_\mu w$, while the quadri-velocity of the observer is $u_\mu=\pa_\mu \tau$, where the coordinate $\tau$ is the time measured by a geodesic observer and the hyper-surfaces of constant $w$ define different past light-cones for different observers.

In this way, the GLC approach allows to express the observed redshift in an exact way
\begin{equation}
1+z=\frac{(k_\mu u^\mu)_s}{(k_\mu u^\mu)_o}=\frac{\Upsilon(\tau_o,w,\tilde\theta^a)}{\Upsilon(\tau_s,w,\tilde\theta^a)}\,,
\label{eq:GLCredshift}
\end{equation}
where the function $\Upsilon$ is evaluated with the same phase $w$ and angles $\tilde{\theta}^a$, but the proper times are different in the numerator and the denominator. This expressions reflects the advantage of the GLC coordinate, describing individual light-cones parametrized by the phase $w$ and the observed angles $\tilde{\theta}^a$.

For our purposes, the significance of Eq. \eqref{eq:GLCredshift} consists of the possibility of expanding $\Upsilon$ up to any order in perturbation theory. This perturbation expansion can be made thanks to the coordinates covariance of the metric tensor
\begin{equation}
g^{\mu\nu}_\text{GLC}=\frac{\pa X^\mu}{\pa x^\alpha}\frac{\pa X^\nu}{\pa x^\beta}\,g^{\alpha\beta}\,,
\label{eq:coordinateTransformation}
\end{equation}
where the GLC coordinate is $X^\mu=\left( \tau,w, \tilde\theta^a \right)$ and a FRW coordinate with metric tensor $g^{\alpha\beta}$ is denoted as $x^\alpha$. Eq. \eqref{eq:coordinateTransformation} can be expanded order by order in terms of the GLC coordinates and metric components as
\begin{align}
X^\mu=\,\bar X^\mu+\delta X^\mu\,,\qquad
\Upsilon=\,\bar\Upsilon+\delta\Upsilon\,,\qquad
\dots&\,\quad
\label{eq:expansion}
\end{align}
For simplicity, we take a cartesian coordinate $x^\alpha =\left( t,x^i \right)$ in FRW and first derive the relation between two coordinates in the background
\begin{equation}
\dot{\bar\tau}=1\qquad,\qquad
\left(a\,\pa_t-n^i\pa_i\right)\bar w=0\,.
\label{eq:background}
\end{equation}
Using the $\tau\tau$ component and the $ww$ one in Eq. \eqref{eq:coordinateTransformation}, we can derive the perturbative relations
\begin{align}
\delta\dot\tau
=&\,\Acal
+\frac{1}{2}\,\delta^{ij}\Bcal_i\Bcal_j
-\frac{1}{2}\Acal^2
+\frac{1}{2\,a^2}\delta^{ij}\pa_i\delta\tau\pa_j\delta\tau
-\frac{1}{a}\delta^{ij}\Bcal_i\pa_j\delta\tau\nonumber\\
=&\,\Acal
-\frac{1}{2}\Acal^2
+\frac{1}{2}\delta^{ij}\left( \Bcal_i-\frac{1}{a}\pa_i\delta\tau \right)\left( \Bcal_j-\frac{1}{a}\pa_j\delta\tau \right)\,,
\label{eq:tautau}
\end{align}
and
\begin{align}
\left( a\,\pa_t
-n^i\pa_i \right)\delta w
=&\,\Acal
-n^in^j\Ccal_{ij}
-n^i\Bcal_i
-\frac{1}{2}\Acal^2
+\frac{1}{2}\,\delta^{ij}\Bcal_i\,\Bcal_j
+2\,\delta^{ij}n^k\Bcal_i\,\Ccal_{jk}
-\delta^{ij}\pa_i\delta w\,\Bcal_j\nonumber\\
&+\frac{1}{2}\left( \delta^{ij}-n^in^j \right)\pa_i\delta w\,\pa_j\delta w
+\Acal\,n^i\pa_i\delta w
-\Acal\,n^in^j\,\Ccal_{ij}\nonumber\\
&-2\,\left( \delta^{ij}-\frac{1}{2}n^in^j \right)n^k\pa_i\delta w\,\Ccal_{jk}
+2\,\,\left( \delta^{jk}-\frac{1}{4}n^jn^k \right)n^i\Ccal_{ij}\,n^l\Ccal_{lk}\,,
\label{eq:ww}
\end{align}
where we have defined $n^i=\frac{x^i}{r}$ and $r=\left( \delta_{ij}x^i\,x^j \right)^{1/2}$. Noting that the differential operator $a\,\pa_t-n^i\pa_i$ is nothing but $-a^{2}\bar k^\mu\pa_\mu$, where $\bar k^\mu=a^{-2}\left(-a,n^i\right)$ is the background quadri-momentum of the photon, the perturbation $\delta w$ to the phase can be expressed at each order as an integration of perturbations along the observer's line of sight. Also note that the combination $n^i\pa_i$ is nothing but the radial derivative $\pa_r$ in polar coordinates so, $n^j\pa_jn^i=\pa_r n^i=0$. Putting it together, Eqs. \eqref{eq:ww} can be written in terms of a polar coordinate $y^i=\left( r,\theta,\phi \right)$ as
\begin{equation}
\delta\dot\tau=\,\Acal
-\frac{1}{2}\Acal^2
+\frac{1}{2}\left( \Bcal_r-\frac{1}{a}\pa_r\delta\tau \right)^2
+\frac{1}{2}\bar\ga^{ab}\left( \Bcal_a-\frac{1}{a}\pa_a\delta\tau \right)\left( \Bcal_b-\frac{1}{a}\pa_b\delta\tau \right)\,,
\label{eq:tautau_radial}
\end{equation}
\begin{align}
\left( a\,\pa_t
-\pa_r \right)\delta w
=&\,\Acal
-\Ccal_{rr}
-\Bcal_r
-\frac{1}{2}\Acal^2
+\frac{1}{2}\,\Bcal^2_r
+\frac{1}{2}\,\bar\ga^{ab}\Bcal_a\,\Bcal_b
+2\,\Bcal_r\,\Ccal_{rr}
+2\,\bar\ga^{ab}\Bcal_a\,\Ccal_{br}\nonumber\\
&-\pa_r\delta w\,\Bcal_r
-\bar\ga^{ab}\pa_a\delta w\,\Bcal_b
+\frac{1}{2}\bar\ga^{ab}\pa_a\delta w\,\pa_b\delta w
+\Acal\,\pa_r\delta w
-\Acal\,\Ccal_{rr}\nonumber\\
&-2\,\bar\ga^{ab}\pa_a\delta w\,\Ccal_{br}
-\pa_r\delta w\,\Ccal_{rr}
+2\,\bar\ga^{ab}\Ccal_{ra}\,\Ccal_{rb}
+\frac{3}{2}\,\Ccal^2_{rr}\equiv\Xi(t,r,\theta,\phi)
\label{eq:ww_radial}\,.
\end{align}
We can then solve Eqs. \eqref{eq:ww} by
\begin{equation}
\delta w(t,r,\theta,\phi)-\delta w_o=-\int_t^{t_o}\frac{dt'}{a(t')}\,\Xi(t',r+\eta(t)-\eta(t'))
=-\int_{\eta(t)}^{\eta_o}d\eta'\,\Xi\left( \eta',\eta_o-\eta' \right)\,,
\label{eq:explicit_w}
\end{equation}
with $\eta(t)=\int_{t_{in}}^t\frac{dt'}{a(t')}$. From this expression, we can easily recover the radial derivative of $\delta w$. Indeed, thanks to the property
\begin{equation}
\frac{d}{d\eta'}\Xi\left( \eta',r+\eta-\eta' \right)=\pa_{\eta'}\Xi\left( \eta',r+\eta-\eta' \right)
-\pa_r\Xi\left( \eta',r+\eta-\eta' \right)\,,
\end{equation}
we derive
\begin{align}
\pa_r\delta w
=&\,\Xi^o_\eta
-\int_\eta^{\eta_o}d\eta'\,\pa_{\eta'}\Xi\left( \eta',r+\eta-\eta' \right)+\pa_r\delta w_o\,,\nonumber\\
a\,\delta \dot w=&\,\Xi_o-\int_\eta^{\eta_o}d\eta'\,\pa_{\eta'}\Xi\left( \eta',r+\eta-\eta' \right)+\pa_r\delta w_o\,.
\label{eq:explicit_drw}
\end{align}
Let us notice that the free function $\delta w_o$ can be function only of $\eta+r$, because we have to satisfy $\left(\pa_\eta-\pa_r\right)\delta w_o=0$. In this way, because $\eta+r=\eta_o$, $\delta w_o$ can be seen as a normalization function at the observer position. We will see in the following subsection how to fix it at linear order, by using the gauge transformation of $\delta w$.

So far, the missing part is the expression for $\Upsilon$. Again, from the coordinates transformation, we consider the $\tau w$ component, i.e.
\begin{align}
\bar\Upsilon=&\,a(t)\nonumber\\
a\,\delta\Upsilon^{-1}=&\,-\Acal
+n^i\left( \Bcal_i+V_i \right)
+a\,\delta\dot w
+\frac{1}{2}\delta^{ij}V_iV_j
-\frac{1}{2}\Bcal^i\,\Bcal_i\nonumber\\
&-a\,\Acal\,\delta\dot w
+\frac{3}{2}\,\Acal^2
-\Acal\,n^i\Bcal_i
+\delta^{ij}\pa_i\delta w\,\left( \Bcal_j+V_j \right)\nonumber\\
&-2\,n^k\Ccal_{ki}\delta^{ij}\left( \Bcal_j+V_j \right)\nonumber\\
=&\,-\Acal
+\left( \Bcal_r+\pa_r\delta\tau \right)
+a\,\delta\dot w
+\frac{1}{2}V_r^2
+\frac{1}{2}\bar\ga^{ab}V_aV_b
-\frac{1}{2}\,\Bcal_r^2
-\frac{1}{2}\bar\ga^{ab}\Bcal_a\,\Bcal_b\nonumber\\
&-a\,\Acal\,\delta\dot w
+\frac{3}{2}\,\Acal^2
-\Acal\,\Bcal_r
+\pa_r\delta w\,\left( \Bcal_r+V_r \right)
+\bar\ga^{ab}\pa_a\delta w\,\left( \Bcal_b+V_b \right)\nonumber\\
&-2\,\left( \Bcal_r+V_r \right)\Ccal_{rr}
-2\,\bar\ga^{ab}\left( \Bcal_a+V_a \right)\Ccal_{br}\nonumber\\
=&\,-\Acal
+\left( \Bcal_r+V_r \right)
+\pa_r\delta w_o+\Xi_o-\int_\eta^{\eta_o}d\eta'\,\pa_{\eta'}\Xi
+\frac{1}{2}V_r^2
+\frac{1}{2}\bar\ga^{ab}V_aV_b
-\frac{1}{2}\,\Bcal_r^2
-\frac{1}{2}\bar\ga^{ab}\Bcal_a\,\Bcal_b\nonumber\\
&+\Acal\,\int_\eta^{\eta_o}d\eta'\,\pa_{\eta'}\Xi
+\frac{3}{2}\,\Acal^2
-\Acal\,\Bcal_r
-\left( \Xi
+\int_\eta^{\eta_o}d\eta'\,\pa_{\eta'}\Xi \right)\,\left( \Bcal_r+V_r \right)\nonumber\\
&-\bar\ga^{ab}\,\left( \Bcal_a+V_a \right)\,\pa_b\int_\eta^{\eta_o}d\eta'\,\Xi
-2\,\left( \Bcal_r+V_r \right)\Ccal_{rr}
-2\,\bar\ga^{ab}\left( \Bcal_a+V_a \right)\Ccal_{br}\nonumber\\
&+\left(\Xi_o+\pa_r\delta w_o\right)\,\left( \Bcal_r+V_r-\Acal\right)\,,
\label{eq:tauw}
\end{align}
where $V_i\equiv\frac{1}{a}\pa_i\delta\tau$. Because of the equality in Eq. \eqref{eq:GLCredshift}, Eqs. \eqref{eq:tauw} allow us to write the observed redshift at second order in perturbation theory. In the following, for a matter of computation, we will focus on $k^\mu u_\mu=\Upsilon^{-1}$. Proving the gauge transformation of this term will automatically provide the gauge transformations of the observed redshift too.

The last part that we want to discuss involves the angles. Just as done so far, linear perturbations of the GLC angles can be obtained by a coordinates transformation as Eq. \eqref{eq:coordinateTransformation}, in particular through its component $w a$
\begin{equation}
a\pa_t\delta\tilde\theta^a
-n^i\pa_i\delta\tilde\theta^a
=\delta^{ij}\pa_i\tilde\theta^a\left(\pa_j \delta w-\Bcal_j-2\,n^k\Ccal_{kj}\right)\,,
\end{equation}
which is solved by
\begin{equation}
\delta\tilde\theta^a=\delta\tilde\theta^a_o-\int_\eta^{\eta_o}d\eta'\left( \bar\ga^{ab}\pa_b\delta w
-\Bcal^a
-2\,\Ccal^{ra}\right)\,.
\label{eq:observed_angles}
\end{equation}
Just as found for $\delta w$, here we have the free function $\delta \tilde\theta^a_o$, which must satisfy the condition $(\pa_\eta-\pa_r)\delta\tilde\theta^a_o=0$. This means that $\delta\tilde\theta^a_o$ can be only function the angle $\tilde\theta^a$ and the combination $\eta+r$, which is exactly the same symmetry allowed by a residual gauge freedom within the GLC gauge \cite{Fleury:2016htl,Mitsou:2017ynv}. This means that the initial condition for the evolution of the angles perturbation $\delta\tilde\theta^a_o$ can be chosen in order to fix the so-called \textit{observational gauge}\footnote{Here we denote by observational gauge the one where $\delta\tilde\theta^a$ is gauge invariant at the observer position, which is meaningful in a perturbative approach.}. We will show this later by using the gauge-transformation properties.

\subsection{Second-order gauge transformations}
\label{App:MT}
The metric perturbations change in a non trivial way under a coordinate transformation, because they are part of a metric tensor but their correspondence to the background depends on the time coordinate only.
Here we provide the gauge transformation for the metric components up to second order in perturbations. We consider a general coordinate transformation with small perturbation $\epsilon^\mu$
\begin{equation}
\tilde x^\mu=x^\mu+\epsilon^\mu
+\frac{1}{2}\epsilon^\rho\pa_\rho\epsilon^\mu+O(3)\,,\qquad\qquad\epsilon^\mu=\left( \epsilon^t,\epsilon^i \right)\,.
\label{eq:coordinates}
\end{equation}
This coordinate transformation generates the gauge transformation of a generic tensor $T$ of any rank
\begin{equation}
T\rightarrow\tilde{T}=T-\mathcal{L}_{\epsilon} T+\frac{1}{2}\mathcal{L}^2_\epsilon T+O(3)\,,
\label{eq:Lie_derivative}
\end{equation}
where $\mathcal{L}_\epsilon$ is the Lie derivative with respect to the vector field $\epsilon^\mu$. Using this relation, we derive the metric perturbations gauge-transform as \cite{Acquaviva:2002ud,Noh:2004bc,Malik:2008im,Yoo:2014sfa}
\begin{align}
\tilde\Acal
=&\,
\Acal-\dot{\epsilon}^t+\frac{1}{2}\left(\epsilon^\rho\partial_\rho\epsilon^t\right)\dot{}
-\epsilon^i\partial_i\Acal
-\left(\epsilon^t\Acal\right)\dot{}
-\Acal\,\dot{\epsilon}^t
+\frac{1}{2}\left(\dot{\epsilon}^t\right)^2
-\frac{a^2}{2}\dot{\epsilon}^i\dot{\epsilon}_i
-a\,\Bcal_i\,\dot{\epsilon}^i\,,\nonumber\\
\tilde \Bcal_i=&\,
\Bcal_i
-\frac{1}{a}\pa_i\left( \epsilon^t-\frac{1}{2}\epsilon^\sigma\partial_\sigma \epsilon^t \right)
+a\dot{\left( \epsilon_i
+\frac{1}{2}\epsilon^\sigma\partial_\sigma \epsilon_i \right)}\nonumber\\
&-\epsilon^t\,\dot\Bcal_i
-\epsilon^k\,\partial_k \Bcal_i
-H\epsilon^t\,\Bcal_i
+\frac{1}{a}\,\dot\epsilon^t\partial_i\epsilon^t
-a\,\dot\epsilon^i\partial_i\epsilon_i\nonumber\\
&-\frac{2}{a}\Acal\,\partial_i\epsilon^t
-\Bcal_j\,\partial_i\epsilon^j
-\Bcal_i\,\dot\epsilon^t
+2\,a\,\Ccal_{ij}\,\dot\epsilon^j
-2\,aH\epsilon^t\,\dot\epsilon_i\,,\nonumber\\
\tilde \Ccal_{ij}=&\,
\Ccal_{ij}
-\left[ \partial_{(i}\epsilon_{j)}-\frac{1}{2}\partial_{(i}(\epsilon^\sigma\partial_\sigma \epsilon_{j)}) \right]
-2H\epsilon^t\Ccal_{ij}
-\epsilon^\rho\partial_\rho \Ccal_{ij}\nonumber\\
&+\delta_{ij}\left[\frac{1}{2}\left( \epsilon^t \right)^2 \left( \dot H+2H^2 \right)
-H\,\left(\epsilon^t-\frac{1}{2}\epsilon^\sigma\partial_\sigma \epsilon^t\right)\right]\nonumber\\
&+\frac{1}{a}\Bcal_{(i}\partial_{j)}\epsilon^t
-2\,\Ccal_{k(i}\partial_{j)}\epsilon^k
+2\,H\epsilon^t \partial_{(i}\epsilon_{j)}
-\frac{1}{2}\partial_i\epsilon^t
\partial_j\epsilon^t
+\frac{1}{2}\partial_i\epsilon^k
\partial_j\epsilon_k\,.
\label{eq:non_linear_gauge_transformations}
\end{align}
Applying Eq. \eqref{eq:Lie_derivative} to the four velocity, we can derive the gauge transformation property of the spatial part as
\begin{equation}
\tilde V_i=\,V_i
+\frac{1}{a}\,\pa_i\left( \epsilon^t-\frac{1}{2}\epsilon^\rho\pa_\rho\epsilon^t \right)
-\epsilon^\rho\pa_\rho V_i
-H\epsilon^t\,V_i
+\frac{1}{a}\Acal\,\pa_i\epsilon^t
-\pa_i\epsilon^j V_j\,.
\label{eq:non_linear_velocity}
\end{equation}
Further decomposition of the metric perturbations in terms of scalar, vector and tensor is possible from Eq. \eqref{eq:Lie_derivative}.

Given the gauge transformation associated with the general coordinate transformation, we can also explicitly derive the gauge transformation of the GLC coordinates. Since the GLC coordinates describe physical quantities at a given spacetime point described by a coordinate $x^\mu$ in FRW, the GLC coordinates will remain invariant under a coordinate transformation. However, with the correspondence to the background fixed, two different coordinates $x^\mu$ and $\tilde x^\mu$ describe two different spacetime points, generating a gauge transformation of the GLC coordinates $X^\alpha$ as
\begin{align}
\tilde{X}^\alpha=&\,X^\alpha-\left( \epsilon^\rho -\frac{1}{2}\epsilon^\sigma\pa_\sigma\epsilon^\rho \right)\pa_\rho X^\alpha+\frac{1}{2}\epsilon^\rho\epsilon^\sigma\pa_\rho\pa_\sigma X^\alpha\,.
\end{align}
We can derive the gauge transformations for $\tau$ and $w$ as
\begin{align}
\widetilde{\delta\tau}=&\,\delta\tau-\epsilon^t+\frac{1}{2}\epsilon^\sigma\pa_\sigma\epsilon^t-\epsilon^\rho\pa_\rho \delta\tau
\label{eq:GTtautau}
\end{align}
and
\begin{align}
\widetilde{\delta w}=&\,\delta w
-\frac{1}{a}\left(\epsilon^t
-\frac{1}{2}\epsilon^\sigma\pa_\sigma\epsilon^t\right)
-n_i\left(\epsilon^i
-\frac{1}{2}\epsilon^\sigma\pa_\sigma\epsilon^i\right)
-\frac{1}{2}\left(\epsilon^t\right)^2\frac{H}{a}
-\epsilon^\rho\pa_\rho \delta w
+\frac{\delta_{ij}-n_i\,n_j}{2\,r}\epsilon^i\epsilon^j\,.
\label{eq:GTww}
\end{align}
Plugging Eqs. \eqref{eq:non_linear_gauge_transformations} and \eqref{eq:non_linear_velocity} into Eqs. \eqref{eq:tautau} and \eqref{eq:ww}, $\phi^{(1)}$ transforms as $-\dot \epsilon^t_{(1)}$. The GT of $\tau$ can be understood as follows: since $\dot\tau^{(1)}=\phi^{(1)}$, the linear order GT is trivially understood. Similarly $\dot\tau^{(2)}$ involves the same terms appearing the integrand of $\delta t^{(2)}_o$, so its gauge transformation, provided in Eq. \eqref{eq:integrand}, is perfectly consistent with Eq. \eqref{eq:GTtautau}. Moreover, the validity of Eqs. \eqref{eq:GTtautau} and \eqref{eq:GTww} is supported also by the fact that they correctly reproduce the gauge transformations of quadri-vectors $u_\mu$ and $k_\mu$ up to second order. Indeed, according to the construction of GLC coordinates, we have that $u_\mu=\pa_\mu\tau$ and $k_\mu=\pa_\mu w$, i.e. $u_\mu$ and $k_\mu$ have no rotational components. This means that $\pa_\rho u_\mu=\pa_\rho\pa_\mu \tau=\pa_\mu\pa_\rho\tau=\pa_\mu u_\rho$ and the same holds for $k_\mu$. With this property, by considering the gradient $\pa_\mu$ of Eqs. \eqref{eq:GTtautau} and \eqref{eq:GTww}, we can show that $\pa_\mu \tau$ and $\pa_\mu w$ transform as covariant vector under gauge transformations, just as expected for $u_\mu$ and $k_\mu$. Moreover, Eq. \eqref{eq:GTww} allows us to fix $\delta w_o$ at linear order. Indeed, from Eq. \eqref{eq:explicit_w}, we get
\begin{align}
\widetilde{\delta w}=&\,-\int_t^{t_o}\frac{dt'}{a(t')}\,\left[ \tilde\Acal-\tilde\Ccal_{rr}-\tilde\Bcal_r \right]+\widetilde{\delta w_o}\nonumber\\
=&\,-\int_t^{t_o}\frac{dt'}{a(t')}\,\left[ \Acal-\Ccal_{rr}-\Bcal_r \right]
+\left[ \frac{\epsilon^t}{a(t)}+n^i\epsilon_i \right]^{t_o}_t+\widetilde{\delta w_o}\,.
\end{align}
In this way, in order to satisfy Eq. \eqref{eq:GTww}, we need that
\begin{equation}
\widetilde{\delta w_o}=\delta w_o-\frac{\epsilon^t_o}{a_o}-\epsilon^r_o\,.
\end{equation}
This relation is satisfied by the normalization
\begin{align}
\delta w_o=&-\frac{\delta t_o}{a_o}-\delta r_o=\int_{t_{in}}^{t_o}dt\left[ \frac{\Acal}{a(t_o)}-\frac{n^i}{a(t)}\left( \Bcal_i+V_i \right) \right]\nonumber\\
=&\,\int_{\eta_{in}}^{\eta_o}d\eta\left[ \frac{a(\eta)}{a(\eta_o)}\Acal-\Bcal_r+V_r \right]\,,
\label{eq:dwo}
\end{align}
which is our choice for the free function $\delta w_o$. Let us comment about this choice of $\delta w_o$. $\delta t_o$ contributes to a monopole in the evaluation of physical quantities so it cannot depend on the angles. On the other hand, $\delta r_o$ is sourced by the radial velocity of the observer along its world-line. It means that it can contribute to the dipole in the physical observables, hence its angular dependence is not null. Then, our normalization implies that $\delta w_o$ can depend on the angles and this choice regards a class of residual gauge freedom within the GLC which is different from the one exploited in \cite{Fleury:2016htl,Mitsou:2017ynv}. In particular, we have that our coordinate system is invariant under the redefinition
\begin{align}
\delta w&\rightarrow \delta w+\delta w_o(\eta+r,\theta^a)\nonumber\\
\delta \tilde\theta^a&\rightarrow\delta \tilde\theta^a+\delta\tilde\theta^a_o(\eta+r,\theta^a)+\pa_b\delta w_o\int^{\eta_o}_{\eta}d\eta'\ga^{ab}(\eta'+r)\,.
\end{align}

The knowledge of $\delta w$ allows us to obtain also the expression for the radial shift at source presented in Eq. \eqref{eq:dtG}. Indeed, we know that
\begin{equation}
w=\eta+r+\delta w=\hat \eta_z+\hat r_z+\frac{\delta t_s}{a}+\delta r_s+\delta w\,
\end{equation}
so, by requiring that $w$ can be identified as the observed past light-cone described by the fiducial coordinates $(\hat \eta_z,\hat r_z)$, i.e. $w=\hat \eta_z+\hat r_z$, we get that
\begin{align}
\delta r_s&=-\delta w-\frac{\delta t_s}{a(\hat t_z)}\nonumber\\
&=\int_t^{t_o}\frac{dt'}{a(t')}\,\left[ \Acal-\Ccal_{rr}-\Bcal_r \right]+\frac{\delta t_o}{a(\hat t_o)}+\delta r_o-\frac{\delta t_s}{a(\hat t_z)}\,,
\label{eq:radial_shift_source}
\end{align}
which is consistent with \cite{Yoo:2014kpa}. Let us also underline that the gauge transformation for $\delta r_z$ is exactly the one required for the radial shift at linear order, namely
\begin{equation}
\widetilde{\delta r_s}=-\widetilde{\delta w}-\frac{\widetilde{\delta t_s}}{a(\hat t_z)}=-\delta w-\frac{\delta t_s}{a(\hat t_z)}+\epsilon^r_s\,,
\end{equation}
just as expected.

In the same way, by explicitly checking the gauge transformation of $\delta\tilde\theta^a$, we can get also the angular shift at the source position, which is crucial for expressing the observables in terms of the observed angles. First of all, let us fix the free-function $\delta \tilde\theta^a_o$ in Eq. \eqref{eq:observed_angles} by studying the gauge-transformation of $\delta \tilde\theta^a$
\begin{equation}
\widetilde{\left(a\pa_t-n^i\pa_i\right)\delta\tilde\theta^a}
=\left(a\pa_t-n^i\pa_i\right)\delta\tilde\theta^a
+\pa_i\tilde\theta^a\left(
-a\dot\epsilon^i
+n^k\pa_{k}\epsilon^i
-\frac{\epsilon^i}{\sqrt{x^kx_k}}\right)\,,
\end{equation}
or, in terms of polar coordinate
\begin{equation}
\widetilde{\left(a\pa_t-\pa_r\right)\delta\tilde\theta^a}
=\left(a\pa_t-\pa_r\right)\left(\delta\tilde\theta^a-\epsilon^a\right)\,.
\end{equation}
Given this, we then impose that $\widetilde{\delta\tilde\theta_o^a}=\delta\tilde\theta_o^a-\epsilon^a_o$, in order to satisfy our requirement that $\widetilde{\delta\tilde\theta^a}=\delta\tilde\theta^a-\epsilon^a_s$. This condition is satisfied if we fix $\delta\tilde\theta^a_o$ to be opposite to the angular part of the spatial shift given in Eq. \eqref{eq:spatial_lapse}
\begin{equation}
\delta\tilde\theta^a_o=-\delta \theta^a_o=-\int_{\eta_{in}}^{\eta_o}\,d\eta\,\left( \Bcal^a+V^a \right)\,.
\end{equation}
Thanks to this choice, we identify $\tilde\theta^a$ as the observed angles so in non linear effects, we have to invert the relation $\tilde\theta^a=\theta^a+\delta\tilde\theta^a$ and expand around $\tilde\theta^a$. Indeed, we can identify $\tilde\theta\equiv \hat \theta^a_s$
\begin{equation}
\tilde\theta^a=\theta^a+\delta\tilde\theta^a=\hat \theta^a_z+\delta\theta^a_s+\delta\tilde\theta^a\,
\end{equation}
which implies
\begin{equation}
\delta\theta^a_s=-\delta\tilde\theta^a=\int_\eta^{\eta_o}d\eta'\left( \bar\ga^{ab}\pa_b\delta w
-\Bcal^a
-2\,\Ccal^{ra}\right)+\delta\theta^a_o\,.
\label{eq:angular_shift_source}
\end{equation}
Again this result is in agreement with the one obtained in \cite{Yoo:2014kpa} and satisfies the required gauge-transformation for the angular shift at the source position. i.e. $\widetilde{\delta\theta^a_s}=\delta\theta^a_s+\epsilon^a_s$

\section{Comparison of the photon wave-vector in two approaches}
\label{Comparison}
We provide the connection between two approaches adopted in this paper. The key quantity is the photon wave-vector $k^\mu$ expressed both in a FRW coordinate and a GLC coordinate (see \cite{Scaccabarozzi:2017ncm} for details). The photon wave-vector in a FRW coordinate can be expressed as
\begin{equation}
k_{FRW}^\mu=\frac{1}{\mathbb{C}\,a^2}\left( -a+\delta\nu,n^i+\delta n^i \right)\,,
\label{eq:geometric_ k}
\end{equation}
where the normalization constant $\mathbb{C}$ is fixed at the observer position with the photon measured energy as $\left( 2\pi\,\mathbb{C}\,a\,\nu \right)_o\equiv 1$. In a GLC coordinate, the photon wave-vector is $k^{GLC}_\mu=\pa_\mu w$, and the transformation from GLC coordinate to a FRW coordinate gives
\begin{equation}
k^{FRW}_\mu=\frac{\pa\,X^\nu}{\pa x_\mu}k^{GLC}_\nu
=\left(\frac{1}{a},n_i\right)+\left(\delta \dot w,\pa_i\delta w\right)\,.
\label{eq:GLC_k}
\end{equation}
By comparing Eqs. \eqref{eq:geometric_ k} and \eqref{eq:GLC_k} we derive
\begin{align}
\delta\nu=&\,2\,a\,\Acal-a\,n^i \Bcal_i-a^2\,\delta\dot w
-a\,\delta n^i\Bcal_i
-2\,\Acal\,\delta\nu\,,\nonumber\\
\delta n^i=&\,-\Bcal^i-2\,\delta^{ik}n^l\,\Ccal_{kl}+\delta^{ij}\pa_j\delta w
+a\,\delta \nu\,\Bcal^i
-2\,\delta^{ij}\Ccal_{jk}\delta n^k\,,
\label{eq:dictionary}
\end{align}
which provides the full vocabulary for the comparison. Eqs. \eqref{eq:dictionary}, combined with Eqs. \eqref{eq:ww}, are sufficient for us to solve the geodesic equations for the wave vector $k^\nu\nabla_\nu k^\mu=0$, i.e.
\begin{align}
\delta\dot\nu=&\,\frac{1}{a}n^i\pa_i\delta\nu+3\,H\,\delta\nu+2\,a\,H\,\delta n^i \,n_i+a\,\delta\Gamma_{tt}^t+\frac{1}{a}\delta\Gamma_{ij}^t\,n^i\,n^j-2\,\delta\Gamma_{ti}^t\,n^i\nonumber\\
&+\frac{1}{a}\delta\nu\,\delta\dot\nu-2\,\frac{H}{a}\,\delta\nu^2+\frac{1}{a}\delta n^i\pa_i\delta\nu+a\,H\,\delta n^i\delta n_i\nonumber\\
&-2\,\delta\Gamma_{tt}^t\delta\nu+2\,\delta\Gamma_{ti}^t\left( \frac{1}{a}n^i\,\delta\nu-\delta n^i \right)+2\,\delta\Gamma_{ij}^t\frac{1}{a^2}\delta n^in^j\,,\nonumber\\
\delta\dot n^i=&\,\frac{1}{a}n^j\pa_j\delta n^i+\frac{1}{a\,r}\left( \delta^i_j-n^i\,n_j \right)\delta n^j+a\,\delta\Gamma_{tt}^i+\frac{1}{a}\delta\Gamma_{jk}^i\,n^j\,n^k-2\,\delta\Gamma_{tj}^i\,n^j\nonumber\\
&-2\,\delta\Gamma_{tt}^i\delta\nu+2\,\delta\Gamma_{tj}^i\left( \frac{1}{a}n^j\,\delta\nu-\delta n^j\right)+2\,\delta\Gamma_{jk}^i\frac{1}{a^2}\delta n^jn^k\nonumber\\
&+\frac{1}{a}\delta \nu\,\delta\dot n^i+\frac{1}{a}\delta n^j\pa_j\delta n^i\,,
\end{align}
where $\delta\Gamma_{\mu\nu}^\rho$ are the perturbed Christoffel symbols
\begin{align}
\delta\Gamma_{tt}^t=&\,\dot\Acal-\dot{(\Acal^2)}+H\,\Bcal^i\Bcal_i+\frac{1}{2}\dot{\left( \Bcal^i\Bcal_i \right)}-\frac{1}{a}\Bcal^i\pa_i\Acal\,,\nonumber\\
\delta\Gamma_{ti}^t=&\,\pa_i\Acal-a\,H\,\Bcal_i-\pa_i\Acal^2
+\frac{1}{2}\Bcal^j\left( \pa_i\Bcal_j-\pa_j\Bcal_i -2\,a\,\dot \Ccal_{ij}\right)+2\,a\,H\,\Acal\,\Bcal_i\,,\nonumber\\
\delta\Gamma_{ij}^t=&\,\dot{\left( a^2\,\Ccal_{ij} \right)}-2\,a^2\,H\,\Acal\,\delta_{ij}+\frac{a}{2}\left( \pa_i\Bcal_j+\pa_j\Bcal_i \right)
+a^2\,H\left( 4\,\Acal^2-\Bcal^k\Bcal_k \right)\delta_{ij}\nonumber\\
&-a\,\Bcal^k\left( \pa_i\Ccal_{jk}+\pa_j\Ccal_{ki}-\pa_k\Ccal_{ij} \right)
-a\,\Acal\left( \pa_i\Bcal_j+\pa_j\Bcal_i \right)-2\,\Acal\dot{\left( a^2\,\Ccal_{ij} \right)}\,,\nonumber\\
\delta\Gamma_{tt}^i=&\,\frac{\pa^i\Acal}{a^2}-\frac{\dot{\left( a\,\Bcal^i \right)}}{a^2}+\frac{\Bcal^i\dot\Acal}{a}-2\,\frac{\delta^{ij}}{a^2}\,\Ccal_{jk}\left[ \pa^k\Acal-\dot{\left( a\,\Bcal^k \right)} \right]\,,\nonumber\\
\delta\Gamma_{tj}^i=&\,\delta^{ik}\dot \Ccal_{kj}+\frac{1}{2\,a}\delta^{ik}\left( \pa_k\Bcal_j-\pa_j\Bcal_k \right)
-\frac{1}{a}\delta^{ik}\Ccal_{kl}\delta^{lm}\left( 2\,a\,\dot \Ccal_{mj} +\pa_m \Bcal_j-\pa_j \Bcal_m \right)\nonumber\\
&+\frac{1}{a}\Bcal^i\left( \pa_j\Acal-a\,H\,\Bcal_j \right)\,,\nonumber\\
\delta\Gamma_{jk}^i=&\,a\,H\,\Bcal^i\delta_{jk}+\delta^{il}\left( \pa_j \Ccal_{kl}+\pa_k \Ccal_{lj}-\pa_l \Ccal_{jk} \right)\nonumber\\
&+\frac{1}{2}\Bcal^i\left( 2\,a\,\dot \Ccal_{jk}-4\,a\,H\,\Acal\,\delta_{jk}+2\,a\,H\,\Ccal_{jk}+\pa_j\Bcal_k+\pa_k\Bcal_j \right)\nonumber\\
&-2\,a\,H\,\delta^{il}\Ccal_{lm}\Bcal^m\,\delta_{jk}
-2\,\Ccal^{il}\left( \pa_j\Ccal_{kl}+\pa_k\Ccal_{lj}-\pa_l\Ccal_{jk} \right)\,.
\end{align}

\bibliographystyle{JHEP}
\bibliography{biblio_timelapse}

\providecommand{\href}[2]{#2}\begingroup\raggedright\begin{thebibliography}{10}

\bibitem{Spergel:2003cb}
{\bf WMAP} Collaboration, D.~N. Spergel et~al., {\it {First year Wilkinson
  Microwave Anisotropy Probe (WMAP) observations: Determination of cosmological
  parameters}},  {\em Astrophys. J. Suppl.} {\bf 148} (2003) 175--194,
  [\href{http://arxiv.org/abs/astro-ph/0302209}{{\tt astro-ph/0302209}}].

\bibitem{Ade:2013zuv}
{\bf Planck} Collaboration, P.~A.~R. Ade et~al., {\it {Planck 2013 results.
  XVI. Cosmological parameters}},  {\em Astron. Astrophys.} {\bf 571} (2014)
  A16, [\href{http://arxiv.org/abs/1303.5076}{{\tt arXiv:1303.5076}}].

\bibitem{Abazajian:2016yjj}
{\bf CMB-S4} Collaboration, K.~N. Abazajian et~al., {\it {CMB-S4 Science Book,
  First Edition}},  \href{http://arxiv.org/abs/1610.02743}{{\tt
  arXiv:1610.02743}}.

\bibitem{York:2000gk}
{\bf SDSS} Collaboration, D.~G. York et~al., {\it {The Sloan Digital Sky
  Survey: Technical Summary}},  {\em Astron. J.} {\bf 120} (2000) 1579--1587,
  [\href{http://arxiv.org/abs/astro-ph/0006396}{{\tt astro-ph/0006396}}].

\bibitem{Colless:2001gk}
{\bf 2DFGRS} Collaboration, M.~Colless et~al., {\it {The 2dF Galaxy Redshift
  Survey: Spectra and redshifts}},  {\em Mon. Not. Roy. Astron. Soc.} {\bf 328}
  (2001) 1039, [\href{http://arxiv.org/abs/astro-ph/0106498}{{\tt
  astro-ph/0106498}}].

\bibitem{Dawson:2012va}
{\bf BOSS} Collaboration, K.~S. Dawson et~al., {\it {The Baryon Oscillation
  Spectroscopic Survey of SDSS-III}},  {\em Astron. J.} {\bf 145} (2013) 10,
  [\href{http://arxiv.org/abs/1208.0022}{{\tt arXiv:1208.0022}}].

\bibitem{Amendola:2012ys}
{\bf Euclid Theory Working Group} Collaboration, L.~Amendola et~al., {\it
  {Cosmology and fundamental physics with the Euclid satellite}},  {\em Living
  Rev. Rel.} {\bf 16} (2013) 6, [\href{http://arxiv.org/abs/1206.1225}{{\tt
  arXiv:1206.1225}}].

\bibitem{Aghamousa:2016zmz}
{\bf DESI} Collaboration, A.~Aghamousa et~al., {\it {The DESI Experiment Part
  I: Science,Targeting, and Survey Design}},
  \href{http://arxiv.org/abs/1611.00036}{{\tt arXiv:1611.00036}}.

\bibitem{Abate:2012za}
{\bf LSST Dark Energy Science} Collaboration, A.~Abate et~al., {\it {Large
  Synoptic Survey Telescope: Dark Energy Science Collaboration}},
  \href{http://arxiv.org/abs/1211.0310}{{\tt arXiv:1211.0310}}.

\bibitem{Yoo:2014sfa}
J.~Yoo and M.~Zaldarriaga, {\it {Beyond the Linear-Order Relativistic Effect in
  Galaxy Clustering: Second-Order Gauge-Invariant Formalism}},  {\em Phys.
  Rev.} {\bf D90} (2014), no.~2 023513,
  [\href{http://arxiv.org/abs/1406.4140}{{\tt arXiv:1406.4140}}].

\bibitem{BenDayan:2012wi}
I.~Ben-Dayan, G.~Marozzi, F.~Nugier, and G.~Veneziano, {\it {The second-order
  luminosity-redshift relation in a generic inhomogeneous cosmology}},  {\em
  JCAP} {\bf 1211} (2012) 045, [\href{http://arxiv.org/abs/1209.4326}{{\tt
  arXiv:1209.4326}}].

\bibitem{Umeh:2014ana}
O.~Umeh, C.~Clarkson, and R.~Maartens, {\it {Nonlinear relativistic corrections
  to cosmological distances, redshift and gravitational lensing magnification.
  II - Derivation}},  {\em Class. Quant. Grav.} {\bf 31} (2014) 205001,
  [\href{http://arxiv.org/abs/1402.1933}{{\tt arXiv:1402.1933}}].

\bibitem{Marozzi:2014kua}
G.~Marozzi, {\it {The luminosity distance-redshift relation up to second order
  in the Poisson gauge with anisotropic stress}},  {\em Class. Quant. Grav.}
  {\bf 32} (2015), no.~4 045004, [\href{http://arxiv.org/abs/1406.1135}{{\tt
  arXiv:1406.1135}}]. [erratum: Class. Quant. Grav.32,179501(2015)].

\bibitem{BenDayan:2012pp}
I.~Ben-Dayan, M.~Gasperini, G.~Marozzi, F.~Nugier, and G.~Veneziano, {\it
  {Backreaction on the luminosity-redshift relation from gauge invariant
  light-cone averaging}},  {\em JCAP} {\bf 1204} (2012) 036,
  [\href{http://arxiv.org/abs/1202.1247}{{\tt arXiv:1202.1247}}].

\bibitem{Fanizza:2013doa}
G.~Fanizza, M.~Gasperini, G.~Marozzi, and G.~Veneziano, {\it {An exact Jacobi
  map in the geodesic light-cone gauge}},  {\em JCAP} {\bf 1311} (2013) 019,
  [\href{http://arxiv.org/abs/1308.4935}{{\tt arXiv:1308.4935}}].

\bibitem{DiDio:2014lka}
E.~Di~Dio, R.~Durrer, G.~Marozzi, and F.~Montanari, {\it {Galaxy number counts
  to second order and their bispectrum}},  {\em JCAP} {\bf 1412} (2014) 017,
  [\href{http://arxiv.org/abs/1407.0376}{{\tt arXiv:1407.0376}}]. [Erratum:
  JCAP1506,no.06,E01(2015)].

\bibitem{Bertacca:2014dra}
D.~Bertacca, R.~Maartens, and C.~Clarkson, {\it {Observed galaxy number counts
  on the lightcone up to second order: I. Main result}},  {\em JCAP} {\bf 1409}
  (2014), no.~09 037, [\href{http://arxiv.org/abs/1405.4403}{{\tt
  arXiv:1405.4403}}].

\bibitem{Bertacca:2014wga}
D.~Bertacca, R.~Maartens, and C.~Clarkson, {\it {Observed galaxy number counts
  on the lightcone up to second order: II. Derivation}},  {\em JCAP} {\bf 1411}
  (2014), no.~11 013, [\href{http://arxiv.org/abs/1406.0319}{{\tt
  arXiv:1406.0319}}].

\bibitem{DiDio:2015bua}
E.~Di~Dio, R.~Durrer, G.~Marozzi, and F.~Montanari, {\it {The bispectrum of
  relativistic galaxy number counts}},  {\em JCAP} {\bf 1601} (2016) 016,
  [\href{http://arxiv.org/abs/1510.04202}{{\tt arXiv:1510.04202}}].

\bibitem{Bohm:2016gzt}
V.~Bohm, M.~Schmittfull, and B.~D. Sherwin, {\it {Bias to CMB lensing
  measurements from the bispectrum of large-scale structure}},  {\em Phys.
  Rev.} {\bf D94} (2016), no.~4 043519,
  [\href{http://arxiv.org/abs/1605.01392}{{\tt arXiv:1605.01392}}].

\bibitem{Pratten:2016dsm}
G.~Pratten and A.~Lewis, {\it {Impact of post-Born lensing on the CMB}},  {\em
  JCAP} {\bf 1608} (2016), no.~08 047,
  [\href{http://arxiv.org/abs/1605.05662}{{\tt arXiv:1605.05662}}].

\bibitem{Marozzi:2016uob}
G.~Marozzi, G.~Fanizza, E.~Di~Dio, and R.~Durrer, {\it {CMB-lensing beyond the
  Born approximation}},  {\em JCAP} {\bf 1609} (2016), no.~09 028,
  [\href{http://arxiv.org/abs/1605.08761}{{\tt arXiv:1605.08761}}].

\bibitem{Lewis:2016tuj}
A.~Lewis and G.~Pratten, {\it {Effect of lensing non-Gaussianity on the CMB
  power spectra}},  {\em JCAP} {\bf 1612} (2016), no.~12 003,
  [\href{http://arxiv.org/abs/1608.01263}{{\tt arXiv:1608.01263}}].

\bibitem{Marozzi:2016qxl}
G.~Marozzi, G.~Fanizza, E.~Di~Dio, and R.~Durrer, {\it {CMB-lensing beyond the
  leading order: temperature and polarization anisotropies}},  {\em Phys. Rev.}
  {\bf D98} (2018), no.~2 023535, [\href{http://arxiv.org/abs/1612.07263}{{\tt
  arXiv:1612.07263}}].

\bibitem{Lewis:2017ans}
A.~Lewis, A.~Hall, and A.~Challinor, {\it {Emission-angle and
  polarization-rotation effects in the lensed CMB}},  {\em JCAP} {\bf 1708}
  (2017), no.~08 023, [\href{http://arxiv.org/abs/1706.02673}{{\tt
  arXiv:1706.02673}}].

\bibitem{Bohm:2018omn}
V.~Bohm, B.~D. Sherwin, J.~Liu, J.~C. Hill, M.~Schmittfull, and T.~Namikawa,
  {\it {On the effect of non-Gaussian lensing deflections on CMB lensing
  measurements}},  \href{http://arxiv.org/abs/1806.01157}{{\tt
  arXiv:1806.01157}}.

\bibitem{Fabbian:2017wfp}
G.~Fabbian, M.~Calabrese, and C.~Carbone, {\it {CMB weak-lensing beyond the
  Born approximation: a numerical approach}},  {\em JCAP} {\bf 1802} (2018),
  no.~02 050, [\href{http://arxiv.org/abs/1702.03317}{{\tt arXiv:1702.03317}}].

\bibitem{Takahashi:2017hjr}
R.~Takahashi, T.~Hamana, M.~Shirasaki, T.~Namikawa, T.~Nishimichi, K.~Osato,
  and K.~Shiroyama, {\it {Full-sky Gravitational Lensing Simulation for
  Large-area Galaxy Surveys and Cosmic Microwave Background Experiments}},
  {\em Astrophys. J.} {\bf 850} (2017), no.~1 24,
  [\href{http://arxiv.org/abs/1706.01472}{{\tt arXiv:1706.01472}}].

\bibitem{Beck:2018wud}
D.~Beck, G.~Fabbian, and J.~Errard, {\it {Lensing Reconstruction in Post-Born
  Cosmic Microwave Background Weak Lensing}},
  \href{http://arxiv.org/abs/1806.01216}{{\tt arXiv:1806.01216}}.

\bibitem{Marozzi:2016und}
G.~Marozzi, G.~Fanizza, E.~Di~Dio, and R.~Durrer, {\it {Impact of
  Next-to-Leading Order Contributions to Cosmic Microwave Background Lensing}},
   {\em Phys. Rev. Lett.} {\bf 118} (2017), no.~21 211301,
  [\href{http://arxiv.org/abs/1612.07650}{{\tt arXiv:1612.07650}}].

\bibitem{Yoo:2017svj}
J.~Yoo and R.~Durrer, {\it {Gauge-Transformation Properties of Cosmological
  Observables and its Application to the Light-Cone Average}},  {\em JCAP} {\bf
  1709} (2017), no.~09 016, [\href{http://arxiv.org/abs/1705.05839}{{\tt
  arXiv:1705.05839}}].

\bibitem{Sachs:1967er}
R.~K. Sachs and A.~M. Wolfe, {\it {Perturbations of a cosmological model and
  angular variations of the microwave background}},  {\em Astrophys. J.} {\bf
  147} (1967) 73--90. [Gen. Rel. Grav.39,1929(2007)].

\bibitem{Fanizza:2015swa}
G.~Fanizza, M.~Gasperini, G.~Marozzi, and G.~Veneziano, {\it {A new approach to
  the propagation of light-like signals in perturbed cosmological
  backgrounds}},  {\em JCAP} {\bf 1508} (2015), no.~08 020,
  [\href{http://arxiv.org/abs/1506.02003}{{\tt arXiv:1506.02003}}].

\bibitem{Giuseppe:2018tim}
F.~Giuseppe, {\it {Relations between physical observables: what is better?}},
  \href{http://arxiv.org/abs/1806.08611}{{\tt arXiv:1806.08611}}.

\bibitem{Yoo:2014vta}
J.~Yoo, {\it {Proper-time hypersurface of nonrelativistic matter flows: Galaxy
  bias in general relativity}},  {\em Phys. Rev.} {\bf D90} (2014), no.~12
  123507, [\href{http://arxiv.org/abs/1408.5137}{{\tt arXiv:1408.5137}}].

\bibitem{Biern:2016kys}
S.~G. Biern and J.~Yoo, {\it {Gauge-Invariance and Infrared Divergences in the
  Luminosity Distance}},  {\em JCAP} {\bf 1704} (2017), no.~04 045,
  [\href{http://arxiv.org/abs/1606.01910}{{\tt arXiv:1606.01910}}].

\bibitem{Cooray:2002mj}
A.~Cooray and W.~Hu, {\it {Second order corrections to weak lensing by large
  scale structure}},  {\em Astrophys. J.} {\bf 574} (2002) 19,
  [\href{http://arxiv.org/abs/astro-ph/0202411}{{\tt astro-ph/0202411}}].

\bibitem{Krause:2009yr}
E.~Krause and C.~M. Hirata, {\it {Weak lensing power spectra for precision
  cosmology: Multiple-deflection, reduced shear and lensing bias corrections}},
   {\em Astron. Astrophys.} {\bf 523} (2010) A28,
  [\href{http://arxiv.org/abs/0910.3786}{{\tt arXiv:0910.3786}}].

\bibitem{Yoo:2014kpa}
J.~Yoo, {\it {Relativistic Effect in Galaxy Clustering}},  {\em Class. Quant.
  Grav.} {\bf 31} (2014) 234001, [\href{http://arxiv.org/abs/1409.3223}{{\tt
  arXiv:1409.3223}}].

\bibitem{Fidler:2015npa}
C.~Fidler, C.~Rampf, T.~Tram, R.~Crittenden, K.~Koyama, and D.~Wands, {\it
  {General relativistic corrections to $N$-body simulations and the Zel'dovich
  approximation}},  {\em Phys. Rev.} {\bf D92} (2015), no.~12 123517,
  [\href{http://arxiv.org/abs/1505.04756}{{\tt arXiv:1505.04756}}].

\bibitem{Fidler:2016tir}
C.~Fidler, T.~Tram, C.~Rampf, R.~Crittenden, K.~Koyama, and D.~Wands, {\it
  {Relativistic Interpretation of Newtonian Simulations for Cosmic Structure
  Formation}},  {\em JCAP} {\bf 1609} (2016), no.~09 031,
  [\href{http://arxiv.org/abs/1606.05588}{{\tt arXiv:1606.05588}}].

\bibitem{Adamek:2017kir}
J.~Adamek, {\it {Perturbed redshifts from N-body simulations}},  {\em Phys.
  Rev.} {\bf D97} (2018), no.~2 021302,
  [\href{http://arxiv.org/abs/1708.07552}{{\tt arXiv:1708.07552}}].

\bibitem{Barausse:2005nf}
E.~Barausse, S.~Matarrese, and A.~Riotto, {\it {The Effect of inhomogeneities
  on the luminosity distance-redshift relation: Is dark energy necessary in a
  perturbed Universe?}},  {\em Phys. Rev.} {\bf D71} (2005) 063537,
  [\href{http://arxiv.org/abs/astro-ph/0501152}{{\tt astro-ph/0501152}}].

\bibitem{Pyne:1995bs}
T.~Pyne and S.~M. Carroll, {\it {Higher order gravitational perturbations of
  the cosmic microwave background}},  {\em Phys. Rev.} {\bf D53} (1996)
  2920--2929, [\href{http://arxiv.org/abs/astro-ph/9510041}{{\tt
  astro-ph/9510041}}].

\bibitem{BenDayan:2013gc}
I.~Ben-Dayan, M.~Gasperini, G.~Marozzi, F.~Nugier, and G.~Veneziano, {\it
  {Average and dispersion of the luminosity-redshift relation in the
  concordance model}},  {\em JCAP} {\bf 1306} (2013) 002,
  [\href{http://arxiv.org/abs/1302.0740}{{\tt arXiv:1302.0740}}].

\bibitem{Scaccabarozzi:2018vux}
F.~Scaccabarozzi, J.~Yoo, and S.~G. Biern, {\it {Galaxy Two-Point Correlation
  Function in General Relativity}},
  \href{http://arxiv.org/abs/1807.09796}{{\tt arXiv:1807.09796}}.

\bibitem{Challinor:2002zh}
A.~Challinor and F.~van Leeuwen, {\it {Peculiar velocity effects in high
  resolution microwave background experiments}},  {\em Phys. Rev.} {\bf D65}
  (2002) 103001, [\href{http://arxiv.org/abs/astro-ph/0112457}{{\tt
  astro-ph/0112457}}].

\bibitem{Scaccabarozzi:2017ncm}
F.~Scaccabarozzi and J.~Yoo, {\it {Light-Cone Observables and Gauge-Invariance
  in the Geodesic Light-Cone Formalism}},  {\em JCAP} {\bf 1706} (2017), no.~06
  007, [\href{http://arxiv.org/abs/1703.08552}{{\tt arXiv:1703.08552}}].

\bibitem{Lewis:2006fu}
A.~Lewis and A.~Challinor, {\it {Weak gravitational lensing of the cmb}},  {\em
  Phys. Rept.} {\bf 429} (2006) 1--65,
  [\href{http://arxiv.org/abs/astro-ph/0601594}{{\tt astro-ph/0601594}}].

\bibitem{RuthBook}
R.~Durrer, {\em The Cosmic Microwave Background}.
\newblock Cambridge University Press, 2008.

\bibitem{Gasperini:2011us}
M.~Gasperini, G.~Marozzi, F.~Nugier, and G.~Veneziano, {\it {Light-cone
  averaging in cosmology: Formalism and applications}},  {\em JCAP} {\bf 1107}
  (2011) 008, [\href{http://arxiv.org/abs/1104.1167}{{\tt arXiv:1104.1167}}].

\bibitem{Fleury:2016htl}
P.~Fleury, F.~Nugier, and G.~Fanizza, {\it {Geodesic-light-cone coordinates and
  the Bianchi I spacetime}},  {\em JCAP} {\bf 1606} (2016), no.~06 008,
  [\href{http://arxiv.org/abs/1602.04461}{{\tt arXiv:1602.04461}}].

\bibitem{Mitsou:2017ynv}
E.~Mitsou, F.~Scaccabarozzi, and G.~Fanizza, {\it {Observed Angles and Geodesic
  Light-Cone Coordinates}},  {\em Class. Quant. Grav.} {\bf 35} (2018), no.~10
  107002, [\href{http://arxiv.org/abs/1712.05675}{{\tt arXiv:1712.05675}}].

\bibitem{Acquaviva:2002ud}
V.~Acquaviva, N.~Bartolo, S.~Matarrese, and A.~Riotto, {\it {Second order
  cosmological perturbations from inflation}},  {\em Nucl. Phys.} {\bf B667}
  (2003) 119--148, [\href{http://arxiv.org/abs/astro-ph/0209156}{{\tt
  astro-ph/0209156}}].

\bibitem{Noh:2004bc}
H.~Noh and J.-c. Hwang, {\it {Second-order perturbations of the Friedmann world
  model}},  {\em Phys. Rev.} {\bf D69} (2004) 104011.

\bibitem{Malik:2008im}
K.~A. Malik and D.~Wands, {\it {Cosmological perturbations}},  {\em Phys.
  Rept.} {\bf 475} (2009) 1--51, [\href{http://arxiv.org/abs/0809.4944}{{\tt
  arXiv:0809.4944}}].

\end{thebibliography}\endgroup

\end{document}